\documentclass[final,5p,times,twocolumn,authoryear]{elsarticle}

\usepackage{amssymb}
\usepackage{lipsum}
\usepackage{tabularx}
\usepackage{booktabs}
\usepackage{stfloats}
\usepackage{xcolor}
\usepackage{aas_macros}
\usepackage{graphicx}
\usepackage{url}
\usepackage{hyperref}
\hypersetup{
    colorlinks=true,
    linkcolor=blue,
    filecolor=magenta,      
    urlcolor=blue,
    citecolor=blue,
}

\journal{Astronomy $\&$ Computing}

\begin{document}

\begin{frontmatter}

\title{Setting SAIL: Leveraging Scientist-AI-Loops for Rigorous Visualization Tools}

\author[CPPM]{Nico Schuster\corref{cor1}}
\cortext[cor1]{Corresponding author. Email address: schuster@cppm.in2p3.fr}

\author[Stewards,UniversityArizona]{Andr\'es N. Salcedo }
\author[Sorbonne]{Simon Bouchard}
\author[USM]{Dennis Frei}
\author[CPPM]{Alice Pisani}
\author[CPPM]{Julian E. Bautista}
\author[CPPM]{Julien Zoubian}
\author[CPPM]{Stephanie Escoffier}
\author[UniversityMichigan,USTC,USTC_SASS]{Wei Liu}
\author[UChicago_Astro,UChicago_KICP]{Georgios Valogiannis}
\author[Sorbonne]{Pauline Zarrouk}

\affiliation[CPPM]{organization={Aix-Marseille Universit\'e, CNRS/IN2P3, CPPM},
            city={Marseille}, 
            country={France}}
            
\affiliation[Stewards]{organization={Department of Astronomy/Steward Observatory, University of Arizona},
            city={Tucson},
            postcode={85721}, 
            state={AZ},
            country={USA}}
\affiliation[UniversityArizona]{organization={Department of Physics, University of Arizona},
            city={Tucson},
            postcode={85721}, 
            state={AZ},
            country={USA}}

\affiliation[Sorbonne]{organization={Sorbonne Universit\'e, CNRS/IN2P3, Laboratoire de Physique Nucl\'eaire et de Hautes Energies (LPNHE)},
            city={Paris},
            postcode={75005}, 
            country={France}}

\affiliation[USM]{organization={Universit\"ats-Sternwarte M\"unchen, Fakult\"at f\"ur Physik, Ludwig-Maximilians Universit\"at},
            city={M\"unchen},
            postcode={81679}, 
            country={Germany}}

\affiliation[UniversityMichigan]{organization={Department of Physics and Leinweber Center for Theoretical Physics, University of Michigan},
            addressline = {450 Church Street},
            city={Ann Arbor},
            postcode={MI 48109}, 
            country={USA}}
            
\affiliation[USTC]{organization={Department of Astronomy, University of Science and Technology of China},
            addressline={96 Jinzhai Road},
            city={Hefei},
            postcode={Anhui 230026},
            country={P.R. China}}

\affiliation[USTC_SASS]{organization={School of Astronomy and Space Sciences, University of Science and Technology of China},
            addressline={96 Jinzhai Road},
            city={Hefei},
            postcode={Anhui 230026},
            country={P.R. China}}

\affiliation[UChicago_Astro]{organization={Department of Astronomy \& Astrophysics, University of Chicago},
            city={Chicago},
            postcode={IL 60637},
            country={USA}}

    \affiliation[UChicago_KICP]{organization={Kavli Institute for Cosmological Physics, University of Chicago},
            city={Chicago},
            postcode={IL 60637},
            country={USA}}

\begin{abstract}

Scientists across all disciplines share a common challenge: the divide between their theoretical knowledge and the specialized skills and time needed to build interactive tools to communicate this expertise. While large language models (LLMs) offer unparalleled acceleration in code generation, they frequently prioritize functional syntax over scientific accuracy, risking visually convincing but scientifically invalid results. This work advocates the Scientist-AI-Loop (SAIL), a framework designed to harness this speed without compromising rigor. By separating domain logic from code syntax, SAIL enables researchers to maintain strict oversight of scientific concepts and constraints while delegating code implementation to AI. We illustrate this approach through two open-source, browser-based astrophysics tools: an interactive gravitational lensing visualization and a large-scale structure formation sandbox, both publicly available. Our methodology condensed development to mere days while maintaining scientific integrity. We specifically address failure modes where AI-generated code neglects phenomenological boundaries or scientific validity. While cautioning that research-grade code requires stringent protocols, we demonstrate through two examples that SAIL provides an effective code generation workflow for outreach, teaching, professional presentations, and early-stage research prototyping. This framework contributes to a foundation for the further development of AI-assisted scientific software.

\end{abstract}


\end{frontmatter}


\section{Introduction}
\label{introduction}

From investigating the behavior of subatomic particles and the mechanics of protein folding, to the chaotic evolution of global weather systems and the formation of structures in our Universe, researchers from all scientific disciplines often share a common challenge. They hold the knowledge required for modeling complex systems, yet they often lack the time and specialized front-end skills to build tools for visualizing their models and concepts. Overcoming this limitation is essential, as real-time, interactive visualizations serve a vital purpose. They can equip scientists with engaging and accessible platforms for public outreach, teaching, and compelling scientific presentations. Moreover, such environments can act as sandboxes for rapidly building intuition about ideas and concepts before committing more resources to further analyses.

The recent evolution of powerful large language models (LLMs) and generative artificial intelligence (generative AI) has initiated a paradigm shift in scientific software development that offers an opportunity to bridge this gap. Specifically, the advent of advanced ``reasoning modes'' allows these models to process multi-step logic and break down complex prompts. While early evaluations on benchmarks such as SWE-bench revealed that initial models could resolve only a fraction of real-world software issues~\citep{Jimenez2024}, modern models have since demonstrated a major improvement in coding proficiency over just two years\footnote{See the official SWE-bench leaderboard: \url{https://www.swebench.com/}}. Applying these tools to research, however, reveals a crucial distinction between software engineering and scientific analysis. Recent studies indicate that although LLMs excel at well-defined coding tasks, they struggle with the open-ended nature of research, often failing to replicate methodologies without human guidance~\citep{Starace2025,HanwenShen2026}. This stems from a fundamental architectural limitation: LLMs are probabilistic models designed to predict statistically likely code sequences, not to apply deductive reasoning to scientific problems~\citep[e.g.,][]{Wang2023,Zahavy2026}. While they offer unprecedented speed, they tend to prioritize functional syntax over scientific rigor. Consequently, they often optimize for computational performance or visual simplicity at the expense of accuracy. When tasked with complex programming, an unguided AI might confidently hallucinate the underlying concepts to ensure functioning code, quietly discarding scientific laws or failing to identify when simplified approximations break down~\citep{Shojaee2025,Song2026}. This risks generating visualizations that appear highly convincing but are fundamentally compromised or invalid.

To mitigate these silent failures and safely harness generative AI in scientific visualization, we advocate for a ``Scientist-AI-Loop'' (\textbf{SAIL}) framework. Adapting the human-in-the-loop model~\citep[e.g.,][]{MosqueiraRey2023} for research, this workflow structurally decouples the science from the syntax, an approach detailed in Section~\ref{Sec:Methodology_SAIL}. Researchers act as the strict domain architects, introducing the necessary concepts and equations, while assigning the AI exclusively to code generation and graphics rendering. The results are standalone HTML-based applications that enable immediate sharing and cross-platform scientific communication. This methodology was derived directly from our experience developing two concrete tools that balance scientific rigor with the necessary approximations required for real-time interactive visualization. To highlight the practical implementation of this approach, Section~\ref{Sec:lensing_visualization} details a real-time Gravitational Lensing rendering, and Section~\ref{Sec:LSS_visualization} introduces a high-performance Large-Scale Structure visualization. Subsequently, Sections~\ref{Sec:Discussion} and~\ref{Sec:Usage} analyze the efficiency, potential pitfalls, and broader educational impact of this methodology. Finally, Section~\ref{Sec:conclusions} summarizes our findings and the framework, emphasizing that while these case studies are rooted in astrophysics, the underlying SAIL framework is universally applicable across all scientific disciplines.

\begin{figure*}[t]
	\centering 
	\includegraphics[width=1.0\textwidth, trim={0.7cm 1.2cm 0.7cm 1.2cm}, clip]{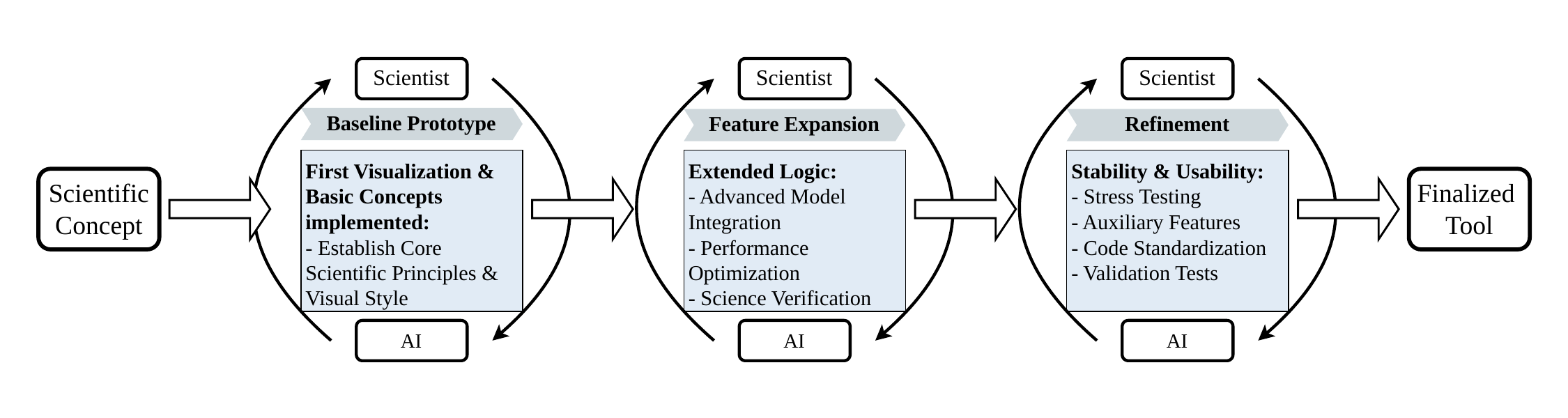}
	\caption{Schematic of the SAIL framework. The workflow transforms a Scientific Concept into a Finalized Tool through three iterative stages: (1) Baseline Prototype for establishing core visuals, (2) Feature Expansion for integrating extended logic and new features, before moving to (3) Refinement for ensuring stability and code standardization. At each stage, the scientist verifies the AI output before proceeding, with the flexibility to cycle back to previous phases to correct errors or add features.} 
	\label{SAIL_framework}
\end{figure*}

\section{Methodology: The Scientist-AI-Loop Framework}
\label{Sec:Methodology_SAIL}

To contextualize the results presented in this work, the visualization tools described in sections~\ref{Sec:lensing_visualization} and~\ref{Sec:LSS_visualization} were developed utilizing state-of-the-art LLMs available at the time of creation. Specifically, we leveraged the advanced reasoning capabilities of Google Gemini Pro 3.0/3.1 and Claude Opus 4.5/4.6 for complex logic and code generation between December 2025 and March 2026. These models were deployed in tandem with integrated AI coding assistants, most notably GitHub Copilot Pro for Researchers. This integration was crucial for creating and managing the complexity of multi-file architectures within the Integrated Development Environment (IDE).

While specific proprietary LLMs were used in the development of examples in subsequent sections, the SAIL framework described here is designed to be model-agnostic. We expect it to be adaptable and effective for a range of generative AI models, although its exact performance may need to be validated first, particularly with open-source alternatives.

\subsection{Charting the Course: The Iterative Workflow}
\label{subsec_phases}

The SAIL framework relies on a continuous, iterative feedback loop between scientists and the employed AI model, condensing the traditional coding-compilation-analysis cycle into almost real-time exchanges. As illustrated in Figure~\ref{SAIL_framework}, this workflow is segmented into three distinct developmental phases, each driven by a continuous exchange between the domain expert and a generative AI model:

\textbf{Phase I: Baseline Prototype}

The foundation of the framework is the \textbf{Scientific Concept} rather than a preexisting software architecture. The scientist's first task is to establish the underlying scientific principles, defining the intended goal (e.g., visualization vs. sandbox tool), and any major constraints. These include not only theoretical limitations, such as conservation of energy or governing equations, but also practical boundaries like realistic runtimes for effective use. The primary goal of this phase is to rapidly generate first visuals from basic concepts, prioritizing a functional prototype over a fully optimized application. Based on these prompted constraints and concepts, the AI generates the initial pipeline. While a single prompt can sometimes suffice, this step typically requires several iterations to refine the first visually correct prototype.

\textbf{Phase II: Feature Expansion}

Once the visual baseline of phase I is validated by the researcher, the workflow shifts to the implementation of \textbf{Extended Logic}. In this phase, the complexity of the underlying scientific principles can be iteratively expanded, or new features can be added. This may involve replacing initial oversimplifications with higher-order approximations, such as the transition from linear models to more advanced non-linear modeling, or the inclusion of additional analysis tools directly within the interactive visualization.

To manage this new sophistication, development proceeds feature-by-feature, where each additional logical component is isolated and tested before the next is added. A central challenge in this phase is maintaining responsiveness while increasing scientific complexity. Simultaneously, the AI is tasked with \textbf{Performance Optimization}. This requires the seamless integration of advancements (e.g., non-linear terms) without destabilizing the existing baseline. Consequently, the expert must continuously verify that the refactored code does not simplify the governing equations for the sake of efficiency. As the codebase expands, a critical risk emerges: the LLM may struggle to retain the full context, inadvertently pruning existing features or essential logic to accommodate context or generation limits. The scientist must therefore treat every major logical expansion as a potential risk, vigilantly checking that new additions do not silently overwrite previously validated work or compromise the integrity of the original scientific concept. To aid in this verification, we recommend fixing random seeds during development to create a deterministic baseline, isolating AI-introduced coding errors from natural statistical variation.

It is worth noting that while the AI can hallucinate concepts or fail in identifying the limits of approximations, it can still serve as an effective consultant. When explicitly prompted by the scientist to suggest standard stabilization techniques or extensions, the LLM can successfully propose relevant concepts. This, however, requires the researcher to first identify the physical breakdown or need for improvement and actively prompt the AI for alternatives, reaffirming the necessity of human oversight.

\textbf{Phase III: Refinement}

The objective of the final stage is to elevate the \textbf{Stability and Usability} of the code. After the rapid expansion of features or logic in Phase II, the focus shifts to the addition of \textbf{Auxiliary Features}. This can include features such as refined user interface (UI) controls, instructional overlays, data export options, or preset scenarios, which are again integrated iteratively, feature-by-feature. Moreover, this stage includes rigorous \textbf{Stress Testing} and \textbf{Validation Tests} to ensure that the code holds up in a variety of conditions and edge cases. Parallel to these functional improvements, the AI can be tasked with \textbf{Code Standardization} (e.g., modularization, refactoring, and documentation) to ensure maintainability, including the creation of unit tests that leverage dedicated functions to systematically verify the correctness of the underlying physics implementations~\citep[e.g.,][]{Chen2023,Pizzorno2024}. This phase is less about adding new physics and more about hardening and reviewing the software, ensuring that the \textbf{Finalized Tool} is intuitive for non-experts and robust against user error, while ensuring scientific fidelity. As a concluding step, we highly encourage accompanying the finalized project with extensive README documentation. Explicitly detailing the underlying concepts and approximations ensures the scientific implementation remains transparent and educational.

\textbf{Continuous Evolution}

Finally, it is important to note that the ``Finalized Tool'' state is not a necessary endpoint. The SAIL workflow is inherently cyclical. Even after a tool is finalized, the framework supports continuous re-entry, where a stable code can serve as the new baseline for subsequent iterations of Feature Expansion or Refinement phases. A researcher might return to Phase II to integrate a newly derived physical term or expand the integrated analysis tools, or choose to update the visualization for a specific audience in Phase III.

\subsection{Steadying the Helm: Architecture and Integrity}
\label{subsec_architecture}

Multi-agent LLM Systems have emerged as powerful methods for scientific workflows, with examples like the ``Virtual Lab'' and ``Denario'' coordinating specialized agents for complex tasks ranging from nanobody design to paper generation across fields~\citep{Swanson2024,Saeedi2025,VillaescusaNavarro2025}. Despite this architectural sophistication, fully autonomous agents are prone to ``misalignment'', where the system loses track of the primary objective or requires frequent human intervention to reset the context~\citep{Cemri2025}.

The SAIL framework addresses these risks through a Human-in-the-Loop partnership. Supported by studies showing that human oversight is critical for maintaining awareness in complex tasks~\citep{Shao2024,Hogg2026}, our methodology casts the AI not as a replacement for the scientist, but as an efficient technical assistant. The scientist retains the role of the ``Domain Architect'', defining the boundary conditions and physical concepts, while the AI operates as the ``Syntax Engine'', responsible for interactive implementation and rendering. This decoupling allows researchers to leverage the rapid coding speed of modern LLMs without sacrificing the scientific verification required for accurate modeling.

\subsection{SAILing from Prototype to Public}
\label{subsec_workflows}

The effective application of SAIL relies on choosing the interaction modality that matches the objective of the application and the developmental phase. We identify two main workflows:

\textbf{1. Single-File Prototyping (Phases I-II):}
The first step is based on interaction with a single model to generate self-contained visualizations, typically embedding HTML, JavaScript, and shaders within a single \texttt{.html} file. This approach minimizes the technical barrier and allows for rapid creation and first validation, as code is created directly in the context window or an interactive canvas. However, as the scope grows, the single-file structure can become difficult to extend, and physics logic might be intertwined with UI handling, decreasing long-term maintainability. Moreover, large files may run into context limits, leading to incomplete or inconsistent outputs. To prevent overwrites, ensure traceability, and counter the lack of automated version control at this stage, we recommend saving files and successful prompts locally after significant alterations, or employing a version control system (e.g., Git) with prompt-related comments.

\textbf{2. Agentic IDE Integration (Phases II-III):}
After an established baseline, it is pertinent to manage the complexity of Feature Expansion. Hence, the workflow transitions to an Agentic IDE context (e.g., Cursor, GitHub Copilot Workspace, Google Antigravity, VS Code) in alignment with professional software engineering best practices. Here, the AI operates as a ``coding agent'', creating a multi-file repository, as well as performing structured tasks such as refactoring and testing. This mirrors emerging multi-agent tools for modern research~\citep{Laverick2024, Swanson2024,Tufano2024,VillaescusaNavarro2025}. While this step imposes a higher entry barrier regarding setup, the resulting codebase is substantially more maintainable and extensible, supporting collaborative development and long-term evolution.

These two workflows are not mutually exclusive, but represent complementary stages of the SAIL framework. The creation of an application typically begins with Single-File Prototyping for rapid conceptual validation. Once validated, the project graduates to Agentic IDE Integration for extension and optimization. For broader use, the verified modules can be ``distilled'' back into lightweight, standalone files, or hosted via web services such as GitHub Pages. This hybrid approach allows researchers to leverage the high-velocity coding capabilities of modern LLMs without sacrificing the fidelity required for public outreach, educational environments, or scientific presentations.

\begin{figure*}[!t]
	\centering 
	\begin{minipage}{0.99\textwidth}
		\centering
		\includegraphics[width=0.495\textwidth]{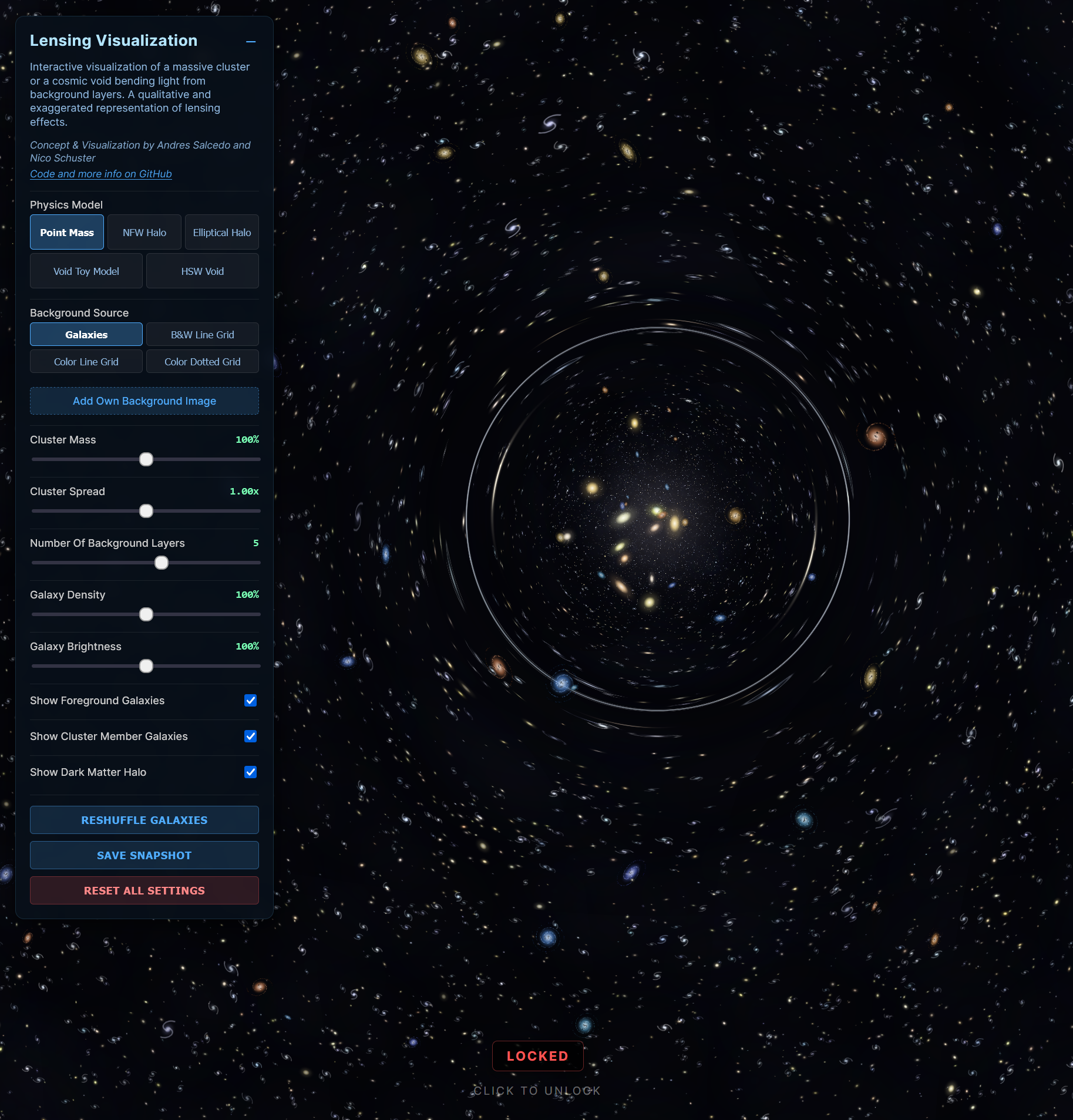}%
        \hfill        
		\includegraphics[width=0.495\textwidth]{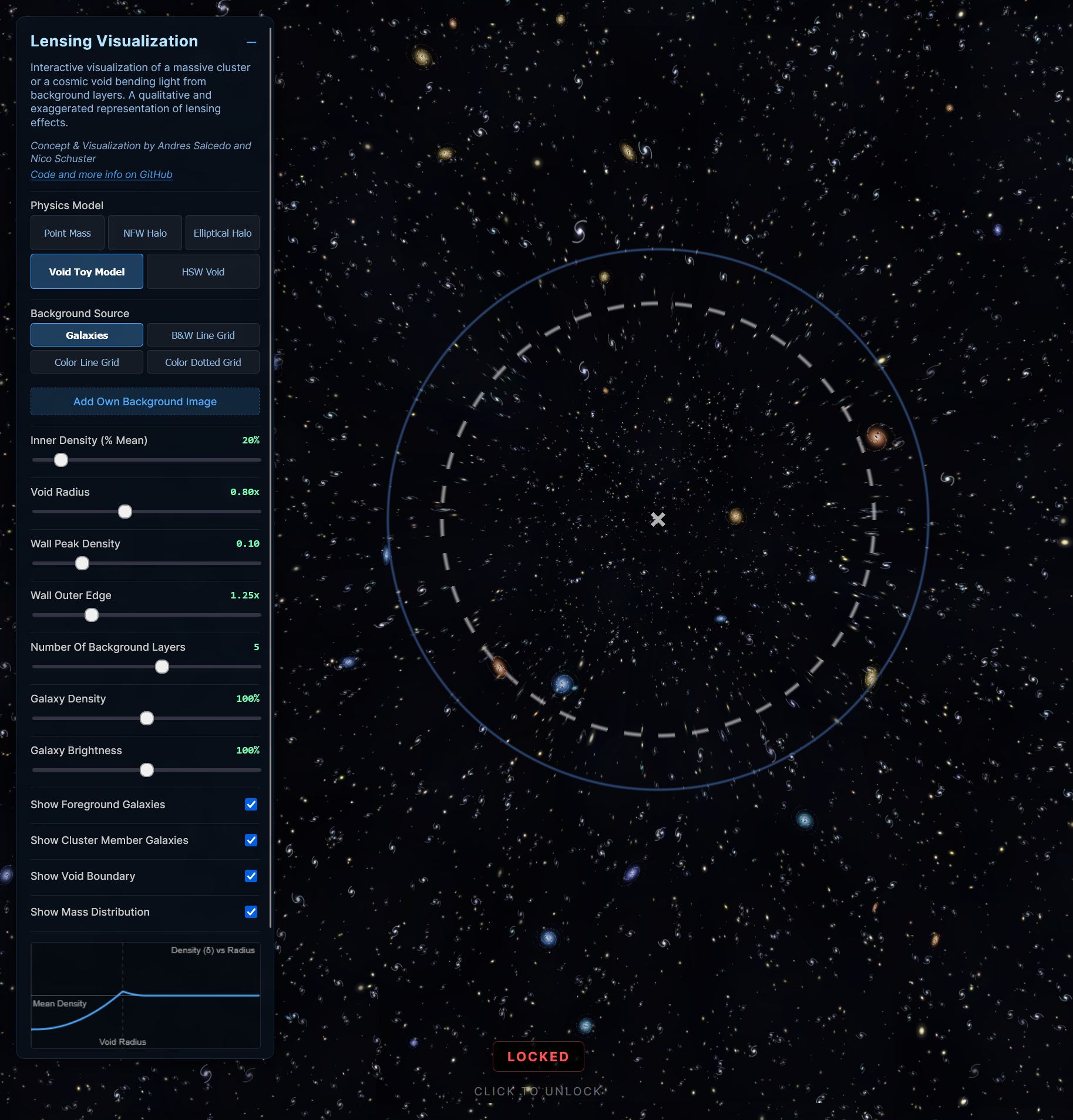}
	\end{minipage}
	\caption{Example snapshots of the Gravitational Lensing Visualization interface for a point mass lens (left) and a cosmic void lens (right). Users can dynamically adjust relative mass/size, lens models (Point Mass, NFW halos, Voids), and other settings to observe exaggerated real-time shear and magnification effects on a multi-plane background. The interactive version is accessible here: \href{https://nicosmo.github.io/lensing_visualization/}{nicosmo.github.io/lensing\_visualization}} 
	\label{fig_lensing}%
\end{figure*}

\section{Case Study I: Gravitational Lensing}
\label{Sec:lensing_visualization}

\textit{Interactive Tool: \href{https://nicosmo.github.io/lensing_visualization/}{Gravitational Lensing Visualization} | Source Code: \href{https://github.com/nicosmo/lensing_visualization/}{GitHub Repository}\footnote{\url{https://github.com/nicosmo/lensing_visualization/}}}

To demonstrate the SAIL methodology in an astrophysical context, we developed a real-time, interactive gravitational lensing module~\citep{schuster_lensing_2026}. This tool allows users to intuitively explore how a variety of mass distributions warp spacetime and distort background light sources, serving as both a pedagogical method for presentations and teaching, as well as a conceptual sandbox for researchers. Figure~\ref{fig_lensing} illustrates the resulting real-time visualization environment.

It is important to note that this tool prioritizes conceptual clarity over absolute physical precision. Rather than using fixed physical units (e.g., solar masses or parsecs/redshift), the visualization employs relative scales to focus on the qualitative behavior of light around the lens. Furthermore, the gravitational lensing effects are purposefully amplified, as weak lensing distortions of individual galaxies are typically on the order of $1\%$ \citep[e.g.,][]{Weinberg2013}. Our exaggeration ensures that subtle effects, such as shear and magnification, are clearly visible within the limited field of view of a browser window. Ultimately, the threshold for required precision is application-dependent and should be determined based on the user's specific research or pedagogical goals.

\subsection{Baseline Implementation}
\label{subsec_lensing_baseline}

The initial Phase I visualization successfully established the baseline physics using standard point mass distributions. The AI efficiently handled the basic deflection angles required for a thin-lens approximation~\citep[e.g.,][]{Bartelmann2001}, implementing an Inverse Ray Shooting technique via custom GLSL fragment shaders~\citep[e.g.,][]{Kayser1986}. This avoids the computationally expensive root-finding required for forward ray tracing while maintaining exact photometric accuracy, allowing the immediate rendering of Einstein rings and shear effects in the browser.

Nevertheless, while the AI immediately excelled at the ``textbook'' case ($M > 0$, point lens), the visual style required significant iterations. The initial output lacked depth and realistic galaxy morphology. To resolve this, we prompted the implementation of representative galaxy types and a multi-plane lensing system. The latter feature treats the background galaxy field as distinct layers with varying depth, introducing parallax effects that provide the user with a 3D intuition of the lensing volume, a feature rarely found in standard static diagrams. To prevent artificial uniformity and increase realism, the implementation included an overlapping size distribution of galaxies for each layer, ensuring variety even as the average angular size decreases with depth. Moreover, to highlight the spacetime distortions caused by the lenses, high-contrast grid lines and dot arrays were implemented as optional background layers, followed by support for custom user-uploaded images. These additions highlight the iterative evolution of the SAIL workflow.

\subsection{Advanced Modeling: NFW Halos and Voids}
\label{subsec_lensing_advanced}

Moving beyond standard point mass approximations required significant intervention in Phase II. We guided the AI to additionally include the mass distribution model of the Navarro-Frenk-White (NFW) profile~\citep{Navarro1997}. The AI successfully translated the 3D density profile into the 2D projected mass distribution needed for the shader logic, demonstrating high proficiency with established models. Following brief refinement in Phase III, a first version of the tool was finalized.

Demonstrating the cyclical nature of the SAIL framework, we subsequently re-entered Phase II to model the demagnifying lensing signal of cosmic voids (convergence $\kappa < 0$). It was here that the most severe friction point emerged. Instead of modeling cosmic voids as local underdensities~\citep{Amendola1999,Krause2013}, the AI’s lack of physical intuition led it to naively invert the NFW profile. This approach produced a visual distortion, but fundamentally mischaracterized the physics by falsely implying that voids are regions of negative mass.

To correct this silent physics failure, we had to step in and provide a valid mass distribution model. For an intuitive mass distribution, we defined a piecewise toy model for the density that consisted of a central under-dense region ($\delta < 0$) surrounded by a dense bounding ridge ($\delta > 0$), before going back to the mean density ($\delta = 0$). This was further refined by our implementation of the commonly used HSW Profile~\citep{Hamaus2014}, a void density profile calibrated to N-body simulations. While physically more accurate, this model proved less intuitive for non-expert users, as the relationship between its parameters and the resulting density shape is difficult for non-experts to predict. The AI was then tasked with implementing these specific models into the shader, ensuring that the visualization was grounded in valid science.

\textbf{Impact and Utility:} The final visualization demonstrates the practical value of the SAIL framework by providing a dynamic, interactive alternative to static diagrams, developed over approximately 4-5 full working days. While primarily pedagogical, the tool’s high fidelity makes it equally suitable for scientific presentations, allowing researchers to demonstrate complex configurations in real-time. Finally, the module’s lightweight, serverless architecture ensures easy accessibility and integration into both lectures and presentations.

\begin{figure*}[t]
	\centering 
	\includegraphics[width=0.99\textwidth]{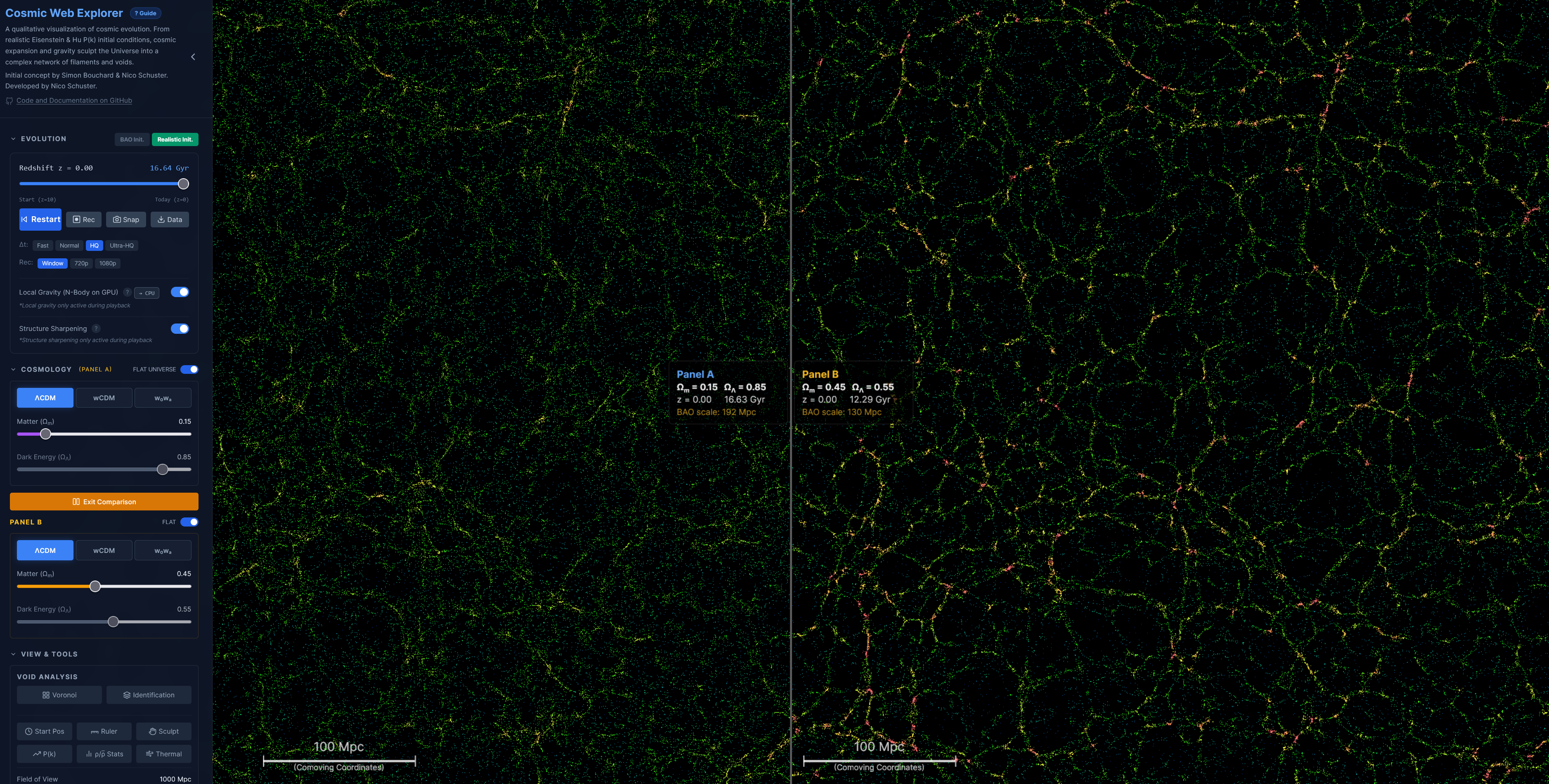}	
	\caption{Example snapshot of the \textit{Cosmic Web Explorer} in ``Cosmology Comparison'' mode. The application allows users to run simplified quasi-N-body visualizations directly within the browser to investigate cosmic structure formation. The initial conditions are generated from a theoretical power spectrum. Additionally, there exists the option for sculpted initial conditions to highlight the BAO signal. In both configurations, the evolution of tracer positions is based on second-order Lagrangian perturbation theory coupled with localized gravity and phenomenological adhesion. The interactive version is accessible here: \href{https://nicosmo.github.io/cosmic_web_explorer/}{nicosmo.github.io/cosmic\_web\_explorer} } 
	\label{fig_LSS}%
\end{figure*}

\section{Case Study II: Cosmic Structure Formation \& BAO}
\label{Sec:LSS_visualization}

\textit{Interactive Tool: \href{https://nicosmo.github.io/cosmic_web_explorer/}{Cosmic Web Explorer Visualization} | Source Code: \href{https://github.com/nicosmo/cosmic_web_explorer/}{GitHub Repository}\footnote{\url{https://github.com/nicosmo/cosmic_web_explorer/}}}

While the previous module visualized fixed optical effects, the second case study focuses on the dynamic formation of cosmic structures~\citep{Schuster_CosmicWebExplorer_2026}. Motivated by recent large-scale survey results~\citep{DESI2025_BAO}, the resulting tool, the \textit{Cosmic Web Explorer}, allows users to dynamically adjust cosmological parameters ($\Omega_\mathrm{m}$, $\Omega_\Lambda$, $w_0$, $w_a$) to observe their impact on the growth of halos, filaments, and voids, specifically highlighting the Baryon Acoustic Oscillation (BAO) signal as a measurable `standard ruler' for the expansion history. It serves as a browser-based interactive sandbox that utilizes simplified methods to simulate the evolution of tens of thousands of mass tracers in a projected 2D universe from high redshift ($z = 10$) to the present day, as depicted in Figure~\ref{fig_LSS}.

To achieve rapid execution speeds, these methods are accelerated by leveraging WebGPU for intense computations. While the AI successfully optimized the performance, the construction of the underlying physics engine highlighted a core friction point that the SAIL framework is designed to solve. The AI proved highly proficient at implementing textbook physics, seamlessly translating linear perturbation theory into code. Conversely, when idealized equations broke down at non-linear scales, the regime where gravity becomes too strong, the process revealed that AI struggles with phenomenological physics. It required constant intervention to integrate appropriate empirical models for more realistic cosmic structures.

\subsection{Baseline Physics: Sculpting the Universe and Gravity}
\label{subsec_LSS_baseline}

The baseline initial conditions (tracer positions at $z = 10$) consist of a background density field, generated from a projected 2D slice of a realistic 3D power spectrum~\citep{Eisenstein1998}. Users can observe the unenhanced evolution of this underlying density field by toggling a secondary mode. However, to ensure visual clarity and facilitate the emergence of the BAO signal at low resolution, the primary visualization required specific tailoring. Because the true BAO signal is statistically weak and difficult to isolate by eye in small volumes or at low particle numbers, we directed the AI to ``sculpt'' a phenomenological initial density field. This was achieved by placing artificial BAO radial shells onto the background, guaranteeing a distinct visual signal throughout the simulation.

To evolve these initial tracer positions, the AI was tasked with implementing linear Lagrangian perturbation theory in the form of the Zel'dovich approximation~\citep{Zeldovich1970}. Furthermore, to drive more realistic clustering, the AI was instructed to implement a ``short-range'' gravity calculation around each tracer. Depending on the user's available CPU or GPU hardware, the AI coded a localized grid to compute 2D $1/r$ Newtonian forces with different ranges. 

While this successfully established the Phase I baseline prototype, the idealized mathematical implementation lacked necessary physical guardrails. As the simulated structures evolved into the non-linear regime, they began exhibiting a variety of unphysical behaviors. Tracers in high-density regions collapsed into singular point masses, while intersecting filaments simply passed through one another without enough physical interaction, a known limitation of Lagrangian models at shell-crossing~\citep{Rampf2021}. These breakdowns perfectly set the stage for the phenomenological interventions required in Phase II of the SAIL workflow.

\subsection{Extended Functionality: Phenomenological Physics}
\label{subsec_LSS_advanced}

To address the unphysical collapses and trajectory crossings identified in the baseline tool, the workflow transitioned into the \textit{Feature Expansion} phase. This stage required the inclusion of phenomenological models similar to those utilized in high-performance N-body simulations~\citep{Hockney1988,Springel2005}. By identifying the specific physical limitations of the linear model and local gravity, we were able to consult the AI for stabilization techniques, such as the introduction of a gravitational softening length ($\epsilon$).

To further improve large-scale evolution, the physics engine was upgraded to Second-Order Lagrangian Perturbation Theory (2LPT)~\citep{Buchert1993}. The AI successfully implemented the required second-order potential, enhancing the formation of filaments and voids. We additionally improved local clustering by implementing an effective gravity strength that scales with cosmological parameters and redshift ($\Omega_m(z)^{0.55}$). This resulted in a more accurate gravity model in comoving coordinates, successfully capturing the suppression of late-time structure growth due to dark energy ($\Omega_\Lambda$).

Lastly, to mitigate the problem of shell-crossing, we prompted the AI agent for potential solutions. The AI suggested and, after our review, implemented a simplified adhesion model. This approach tracks the number of trajectory crossings for each particle and damps their velocity accordingly. The specific choice of parameter values was dictated entirely by our feedback, ensuring the visual output remained physically realistic. This phase highlighted the iterative nature of SAIL: the researcher identifies the need for improvement, the AI proposes and implements the syntax for the solution, and the researcher verifies the result.

\subsection{Refinement: Cosmology Comparison, 2PCF, and Voids}
\label{subsec_LSS_refined}

\begin{figure}[t]
	\centering
	\includegraphics[width=0.98\columnwidth]{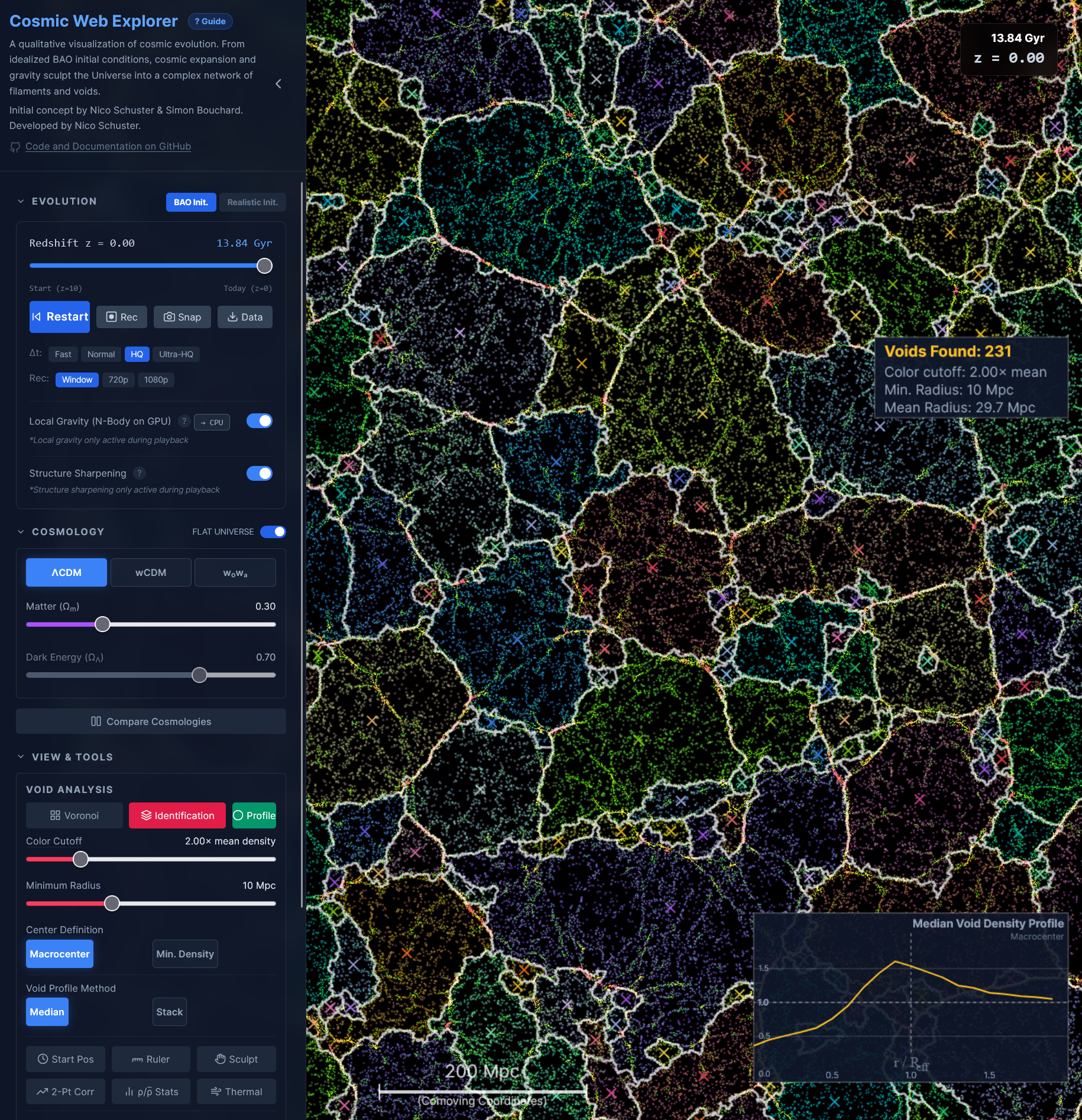}	
	\caption{Snapshot of the \textit{Cosmic Web Explorer} in void-analysis mode, with BAO idealized initial conditions. This module demonstrates the evacuation of matter from underdense regions into the surrounding filaments and nodes. The tool facilitates the study of void geometry and density through interactive merging and density profiles, providing a visual intuition for the volume-dominating components of large-scale structure.}
	\label{fig_voids}%
\end{figure}

Reflecting the cyclical and non-linear nature of the SAIL framework, the development of additional features was not strictly reserved until after completion of the physics engine. Several core components like the void analysis were already functioning before the transition to 2LPT, demonstrating that the SAIL framework is not rigidly sequential, allowing researchers to add functionality even to baseline models.

To maximize its utility, we implemented a split-screen Cosmology Comparison Mode for simultaneous analysis of structure formation across varying parameters. To further support the creation of materials for scientific presentations and outreach, we instructed the AI to integrate a video export function that saves the canvas into standard formats. By recording one frame per computational time step, this bypasses the visual lag typically caused by high computational loads during real-time screen recording.

Beyond visual comparison, the tool was expanded to include real-time quantitative analysis. We instructed the AI to compute a simplified two-point correlation function (2PCF), correlating known BAO centers directly with surrounding tracers rather than computing all tracer pairs, to clearly highlight the evolution of the measured signal.

Finally, because the extended underdense regions of the cosmic web serve as pristine cosmological probes for dark energy, the void identification module (see Figure~\ref{fig_voids}) was designed to mirror the topological watershed methodologies of \textsc{vide}~\citep{Sutter2015} and \textsc{zobov}~\citep{Neyrinck2008}, once more illustrating the necessity of expert intervention.  Because the \textit{Cosmic Web Explorer} operates at a much lower tracer resolution than standard N-body simulations, we had to intervene and prompt the AI to implement additional watershed criteria not present in the standard \textsc{vide} pipeline. To prevent excessive merging, we restricted the grouping of adjacent regions to those separated by sufficiently broad and underdense boundaries. We then added an interactive slider, allowing users to optionally merge the remaining small voids into adjacent larger neighbors. Adjusting this slider qualitatively reproduces how merging impacts the radial density profiles in high-resolution 3D cosmological simulations~\citep{Schuster2023}. Additionally, by alternating the center definition between the geometric barycenter and the minimum density center, the tool reproduces the distinct profile characteristics of voids from \textsc{vide}~\citep[e.g.,][]{Sutter2015} and \textsc{revolver}~\citep[e.g.,][]{Nadathur2020}, respectively. By successfully reproducing these qualitative topological metrics within a browser environment, the SAIL framework demonstrates how generative AI can be safely harnessed to create highly accessible, interactive presentation applications for complex fields of active research.

\textbf{Impact and Utility:} The resulting web application, developed in less than 80 hours, highlights the SAIL framework’s effectiveness in translating elaborate concepts into a real-time browser experience. Its phenomenological fidelity in visualizing large-scale structure formation and its analysis options make it an ideal pedagogical and presentation tool for both educational and professional scientific settings. To further highlight the flexibility of the SAIL framework, a complementary visualization with a dedicated focus on the evolution of BAO rings alongside the expansion of the Universe was also developed~\footnote{\url{https://sbouchard.fr/bao-visu/gemini-v1/}}. This parallel tool demonstrates how the same overarching methodology can yield diverse implementations and representations tailored to specific scientific or pedagogical goals.

\section{Discussion: Efficiency, Integrity, and Universality}
\label{Sec:Discussion}

\subsection{Accelerated Development and Conceptual Oversight}
\label{subsec_acceleration}

Although precise time-tracking was not conducted during the development of the presented case studies, the quantitative shift in development speed is striking. Developing robust, real-time interactive environments would typically demand months of dedicated software engineering. In contrast, the SAIL workflow drastically condenses this timeline, requiring approximately 4-5 full working days for the Gravitational Lensing module and less than 80 hours for the cosmic structure formation web application. This efficiency directly addresses the technical barrier identified in the introduction, empowering researchers to build engaging applications without needing to acquire specialized front-end or graphics programming skills. Such a streamlined workflow is made even more appealing by how quickly a working model can be established, with the core of the functionality coming together in a minority of the total time. Considering the extensive capabilities embedded in the final modules, scientists leveraging this framework can create basic, functional versions of complex systems with very little initial time commitment.

This acceleration stems from the SAIL framework's core principle: structurally decoupling scientific logic from computational syntax. While the general use of AI naturally accelerates code generation, SAIL leverages this by shifting the researcher's focus toward conceptual oversight. By designating the AI as the syntax generator, the framework positions the researcher to act as a high-level manager throughout the iterative loop. Because the AI optimizes for functional code rather than scientific truth, the scientist’s time is redirected from manually implementing features and fixing syntax bugs to verifying that explicit equations are correct. Moreover, this allows the scientist to focus on their physical intuition, actively identifying where idealized models fail and introducing phenomenological adjustments for achieving realistic visual fidelity.

\begin{table*}[!t]
\centering
\caption{Analysis of AI Hallucinations and Required Scientist Interventions. The failure categories and examples presented are drawn directly from the case studies (Sections~\ref{Sec:lensing_visualization} and~\ref{Sec:LSS_visualization}), illustrating the progression of errors from physical inaccuracies to architectural instability.}
\label{tab_AI_failures}
\renewcommand{\arraystretch}{1.4}
\begin{tabularx}{\textwidth}{@{} >{\raggedright\arraybackslash}p{2.8cm} X X X @{}}
\toprule
\textbf{Failure Category} & \textbf{Encountered Symptom} & \textbf{Root Cause} & \textbf{SAIL Resolution} \\ 
\midrule
\textbf{1. Domain Knowledge Deficit} & 
Modeled negative convergence ($\kappa < 0$) incorrectly inside cosmic voids, implying the existence of negative mass. & 
AI applied a generic density inversion instead of realistic density distributions. & 
The scientist manually defined the correct piecewise void density profile. \\ 
\addlinespace
\textbf{2. Contextual Misapplication} & 
Unphysical particle-crossing: tracers passed directly through each other instead of forming stable filaments. & 
AI correctly applied the theoretical model but failed to recognize the breakdown of these equations in the non-linear regime. & 
The scientist instructed the AI to implement a phenomenological ``adhesion model'' to simulate more realistic structure formation. \\
\addlinespace
\textbf{3. Silent Logic Overwrite} & 
Altered visuals: AI removed code blocks that accounted for a correct calculation of densities, thereby changing the visual output. & 
AI refactored the physics loop during WebGPU optimization, quietly deleting the previously validated visualizations. & 
Transitioned to an Agentic IDE workflow. Explicitly instructed the AI to preserve unrelated code during functionality updates. \\
\addlinespace
\textbf{4. Architectural Instability} & 
Intertwined codebase: Interface elements breaking when functionality was updated, preventing feature expansion. & 
AI placed rendering logic, UI, and complex math within the same file, losing context as token limits were reached. & 
The scientist enforced modularity, separating the physics engine from the graphics pipeline into independent, maintainable files. \\
\bottomrule
\end{tabularx}
\end{table*}

\subsection{Generative Pitfalls and Code Maintenance}
\label{subsec_pitfalls}

The critical need for conceptual oversight in SAIL is best illustrated by the specific errors encountered during development. When left unguided in complex scientific modeling, AI coding assistants are prone to generating inaccurate models or breaking validated functional code as the project scales. To formalize the challenges overcome by the workflow, Table~\ref{tab_AI_failures} classifies the typical AI failures observed across our case studies detailed in Sections~\ref{Sec:lensing_visualization} and~\ref{Sec:LSS_visualization}. These range from fundamental gaps in domain knowledge to the silent degradation of the software architecture.

As demonstrated in Table~\ref{tab_AI_failures}, the nature of AI failures shifts as the code base increases. Early development is typically characterized by scientific knowledge deficits or breakdowns (Categories 1 and 2), which can be solved through targeted prompting. As the functionality expands, however, these errors transition into architectural instability and silent logic overwrites (Categories 3 and 4), often due to reaching token limits. To mitigate or reduce such failures, the SAIL framework suggests transitioning to the Agentic IDE workflow in the later stages of development. Operating in such environments allows the researcher to enforce multi-file management, isolating the scientific and visualization engines from other logic to increase long-term maintainability and prevent overwriting verified logic (see Section~\ref{subsec_workflows}).

\subsection{Scalability Across Scientific Fields}
\label{subsec_scalability}

While the case studies presented in this work are rooted in astrophysics, the SAIL framework addresses a common bottleneck in most modern sciences: the disconnect between domain knowledge and specialized software implementation skill, such as developing interactive web visualizations using JavaScript and HTML. Furthermore, the tendency of LLMs and generative AI to prioritize functional code over scientific accuracy is not unique to cosmology, representing a challenge faced across all scientific domains.

Even though SAIL was only tested on two astrophysics-focused examples, the core idea of letting the scientist oversee scientific logic while the AI handles the code is expected to be easily adaptable to other fields, from the microscopic to the macroscopic. At the smallest scales, particle physicists could potentially use SAIL to build interactive visualizations of collider events. In biochemistry, a researcher could guide the AI to create rendering pipelines for protein folding. Moving to planetary scales, climate scientists could use the SAIL framework to visualize shifting ocean currents, stepping in specifically to correct the dynamic boundaries that AI often fails to recognize. Beyond the natural sciences, fields like economics could use this framework to generate visualizations of supply chain dynamics. Finally, returning to astrophysics, the framework could help visualize phenomena ranging from supernova light curves and their standardization to binary black hole or neutron star mergers and the resulting gravitational waves. Admittedly, the exact use across diverse fields and applications should be assessed on a case-by-case basis.

Beyond these illustrative examples, SAIL offers a generalized pathway to produce high-fidelity, interactive modules by focusing the scientist's cognitive load to conceptual oversight. It allows experts to easily build reliable, scientifically accurate applications, simplifying the development of interactive software for both research and education.

\section{Interactive Scientific Communication and Application}
\label{Sec:Usage}

Until recently, the time required to code interactive visualizations restricted their use to broad, general-purpose educational software. Investing months of development into a highly specialized web application was largely impractical or unattainable for researchers. By condensing development time from months to mere days (as discussed in Section~\ref{subsec_acceleration}), the SAIL framework resolves this bottleneck. Researchers can now efficiently create and scale the complexity of their applications to suit specific audiences. This enables an adaptable spectrum of interactive tools, ranging from simplified, high-level modules for public education to highly niche, limited-purpose sandboxes for expert collaboration.

\subsection{Modules for Outreach and Higher Education}
\label{subsec_outreach}

Interactive visualizations offer pedagogical advantages for both public outreach and university-level education. For general audiences, simplified configurations can intuitively illustrate complex scientific principles. By allowing the public to dynamically adjust underlying system parameters that represent key variables in active areas of research, the software translates abstract concepts into observable phenomena. Within this context, high visual fidelity functions as a ``gamified'' learning mechanism rather than a mere cosmetic enhancement. The ability to instantly observe a system react to user inputs supports an intuitive engagement that static media cannot provide.

In university classrooms, the depth of such applications can be precisely tuned to match the students' expertise. This is particularly valuable in advanced Master's-level courses, which are traditionally dominated by more theoretical lectures. Here, interactive modules translate theoretical derivations into observable outcomes. By altering system parameters in real time, students can directly observe how changes in governing equations affect the behavior of the simulated model, transforming theoretical concepts into an active exploration of scientific mechanics and further developing scientific intuition.

\subsection{Scientific Presentations and Sandbox Prototyping}
\label{subsec_presentations}

For scientific presentations, interactive visuals consistently engage audiences more effectively than traditional static plots. Researchers can leverage this fact to create highly targeted visual aids with minimal effort. The SAIL framework allows scientists to build accessible modules that introduce fundamental principles to non-experts, or alternatively, to engineer highly niche visualizations that explain specific concepts to peers. By replacing static plots with dynamic translations of new theoretical developments, these custom tools facilitate crucial conversations between theoretical modelers and observational experimentalists. Because these tools run in lightweight browser environments, researchers can share complex interactive models via a simple web link or single file, making them instantly accessible to peers without any installation or setup.

While the major focus of this methodology is enhancing scientific communication, a valuable secondary application is its potential as a private theoretical sandbox. Theorists can rapidly visualize new frameworks and phenomenological models directly within a web browser before committing extensive effort to full-scale computational analyses.

Finally, the utility of this interactive approach is supported by qualitative feedback from experts who evaluated the scientific fidelity of our case studies. Notably, many users immediately identified the tools' strong potential for communication, expressing a desire to integrate the modules into their own lectures and conference talks. They specifically highlighted the intuitive nature of real-time parameter manipulation and the clarity provided by high visual fidelity, confirming that the SAIL workflow successfully enables elevated scientific exchange.

\section{Summary and Conclusions}
\label{Sec:conclusions}

Scientists across all disciplines often face a persistent bottleneck: translating deep domain knowledge into interactive visual tools for outreach and peer presentations demands significant investments of time and specialized programming skills that directly compete with their active research. While the advent of large language models offers an unparalleled acceleration in software generation, these tools fundamentally optimize for executable code rather than rigorous scientific accuracy, risking the production of visually convincing but scientifically flawed results. To mitigate these inherent hazards, this work presented the ``Scientist-AI-Loop'' (SAIL) framework. By structurally decoupling the science from code creation, SAIL frames the researcher as the conceptual architect while the AI serves strictly as the syntax engine, with the former ensuring that scientific rigor and visual fidelity drive the final product.

The core of this methodology is a continuous loop of AI implementation and scientific verification. The framework's practical utility was demonstrated via two distinct astrophysical case studies. Through these applications, the translation of physical models into interactive web applications was condensed to a timescale of mere days, all while maintaining scientific fidelity. Although these examples are focused on cosmology, the underlying separation of domain logic from coding syntax in this adaptation of the human-in-the-loop model is expected to be applicable to a wide range of both scientific and non-scientific disciplines. Additionally, SAIL offers a flexible platform for sandboxing theoretical ideas and testing scientific intuition before committing to more resource-intensive research phases. While the principles of the SAIL workflow can be extended to broader scientific computing and data analysis, such applications require even more rigorous verification protocols, as the standards for research-grade coding demand a level of precision that exceeds that of pedagogical visualization. Nevertheless, by minimizing the traditional coding bottleneck, SAIL enables researchers to rapidly and independently build interactive tools, fundamentally transforming how scientific concepts can be shared in public outreach, university classrooms, and professional peer-to-peer presentations. However, generating such visualizations is only the first step. Their actual impact relies on scientists promoting and integrating them to reach their intended audience. Ultimately, the SAIL framework supports scientists in fully realizing the fundamental purpose of science itself: advancing the frontier of human knowledge and making those discoveries accessible to the wider world.

\section*{Practical Implementation Guide}

To transition the SAIL framework from a conceptual model into a reproducible practice, we outline a summary based on the structured protocols and experiences described in this work. The workflow begins by establishing a \textbf{deterministic baseline} from initial scientific concepts. This requires fixing random seeds to guarantee consistent outputs during development and clearly defining \textbf{scientific constraints}, as well as realistic runtimes, within the initial prompts. During the early single-file prototyping stage (Phases I and II), we recommend maintaining a ``prompt-to-output'' log. By saving functional code versions alongside the specific prompts that generated them, either locally or via a version control system (e.g., Git) with prompt-annotated commits, researchers \textbf{ensure traceability} and a reliable method for reverting to previous versions if the model hallucinates or removes logic.

As the project scales and transitions into an agentic IDE environment (between Phases II and III), development should proceed strictly \textbf{feature-by-feature}. This modular approach allows for review checkpoints after every code generation to verify that the AI has not silently overwritten previously validated physics or logic to accommodate new features or context limits. This verification can be further hardened by tasking the AI with \textbf{rigorous stress testing of edge cases} and the creation of unit tests that systematically check the underlying implementations. Crucially, any automated refactoring must be reviewed to ensure mathematical rigor is maintained during code optimization. The process culminates in \textbf{code standardization} by outsourcing specific logic into separate files, and the creation of extensive README documentation. By explicitly detailing the generative models utilized, the specific physical approximations applied, and the prompt history, the scientist ensures that the finalized tool remains a transparent, scientifically rigorous, and educational environment.

\section*{Software and Code Availability}
\label{data}

The interactive tools and source code developed for the case studies presented in this work are publicly available. The Gravitational Lensing Visualization (Section~\ref{Sec:lensing_visualization}) can be accessed at \url{https://nicosmo.github.io/lensing_visualization/} and \url{https://github.com/nicosmo/lensing_visualization/}. Similarly, the Cosmic Structure Formation module (Section~\ref{Sec:LSS_visualization}) and its underlying code are hosted at \url{https://nicosmo.github.io/cosmic_web_explorer/} and \url{https://github.com/nicosmo/cosmic_web_explorer/}. Additionally, the complementary BAO visualization is available at \url{https://sbouchard.fr/bao-visu/gemini-v1/}.

To ensure independently developed tools reach the widest possible audience, we aim to curate a community-driven web directory. We invite researchers to reach out and share the links to their own tools so we can feature them in a growing collection.

\section*{Acknowledgements}
We thank Pierre Boccard, Marie-Claude Cousinou, Jahmour J. Givans, Steffen Hagstotz, Nico Hamaus, Geray Karademir, Juan Mena-Fern\'{a}ndez, Lucas Sauniere, and Leander Thiele for discussions and feedback on the interactive visualization tools and earlier drafts of this manuscript. N.S. would like to thank Tim Eifler, Elisabeth Krause, and Enrique Paillas for their hospitality at the CosmoLab of the University of Arizona, which facilitated the discussions that led to this project. A.P. and N.S. acknowledge support from the french government under the France 2030 investment plan, as part of the Initiative d’Excellence d’Aix-Marseille Universit\'e - A*MIDEX AMX-22-CEI-03. A.P. acknowledges support from the European Research Council (ERC) under the European Union's Horizon programme (COSMOBEST ERC funded project, grant agreement 101078174). The project leading to this publication has received funding from Excellence Initiative of Aix-Marseille University - A*MIDEX, a French ``Investissements d'Avenir'' program (AMX-20-CE-02 - DARKUNI). G.V. acknowledges the support of the Eric and Wendy Schmidt AI in Science Fellowship at the University of Chicago, a program of Schmidt Sciences. The algorithm used to compute and render the caustic and critical curves of the Gravitational Lensing Visualization is built on formalisms from the lenstronomy~\footnote{\url{https://github.com/lenstronomy/lenstronomy}} Python package~\citep{Birrer2015,Birrer2018,Birrer2021}. Furthermore, we thank GitHub for supporting academic research by providing access to GitHub Copilot Pro, which was instrumental in generating the codebase for the interactive web visualizations and testing the framework's methodologies.

\bibliographystyle{elsarticle-harv} 
\bibliography{bibmain}

\begin{thebibliography}{41}
\expandafter\ifx\csname natexlab\endcsname\relax\def\natexlab#1{#1}\fi
\providecommand{\url}[1]{\texttt{#1}}
\providecommand{\href}[2]{#2}
\providecommand{\path}[1]{#1}
\providecommand{\DOIprefix}{doi:}
\providecommand{\ArXivprefix}{arXiv:}
\providecommand{\URLprefix}{URL: }
\providecommand{\Pubmedprefix}{pmid:}
\providecommand{\doi}[1]{\href{http://dx.doi.org/#1}{\path{#1}}}
\providecommand{\Pubmed}[1]{\href{pmid:#1}{\path{#1}}}
\providecommand{\bibinfo}[2]{#2}
\ifx\xfnm\relax \def\xfnm[#1]{\unskip,\space#1}\fi
\bibitem[{{Adame} et~al.(2025){Adame}, {Aguilar}, {Ahlen}, {Alam}, {Alexander}, {Alvarez}, {Alves}, {Anand}, {Andrade}, {Armengaud}, {Avila}, {Aviles}, {Awan}, {Bahr-Kalus}, {Bailey}, {Baltay}, {Bault}, {Behera}, {BenZvi}, {Bera}, {Beutler}, {Bianchi}, {Blake}, {Blum}, {Brieden}, {Brodzeller}, {Brooks}, {Buckley-Geer}, {Burtin}, {Calderon}, {Canning}, {Carnero Rosell}, {Cereskaite}, {Cervantes-Cota}, {Chabanier}, {Chaussidon}, {Chaves-Montero}, {Chen}, {Chen}, {Claybaugh}, {Cole}, {Cuceu}, {Davis}, {Dawson}, {de la Macorra}, {de Mattia}, {Deiosso}, {Dey}, {Dey}, {Ding}, {Doel}, {Edelstein}, {Eftekharzadeh}, {Eisenstein}, {Elliott}, {Fagrelius}, {Fanning}, {Ferraro}, {Ereza}, {Findlay}, {Flaugher}, {Font-Ribera}, {Forero-S{\'a}nchez}, {Forero-Romero}, {Frenk}, {Garcia-Quintero}, {Gazta{\~n}aga}, {Gil-Mar{\'\i}n}, {Gontcho a Gontcho}, {Gonzalez-Morales}, {Gonzalez-Perez}, {Gordon}, {Green}, {Gruen}, {Gsponer}, {Gutierrez}, {Guy}, {Hadzhiyska}, {Hahn}, {Hanif}, {Herrera-Alcantar}, {Honscheid}, {Howlett},
  {Huterer}, {Ir{\v{s}}i{\v{c}}}, {Ishak}, {Juneau}, {Kara{\c{c}}ayl{\i}}, {Kehoe}, {Kent}, {Kirkby}, {Kremin}, {Krolewski}, {Lai}, {Lan}, {Landriau}, {Lang}, {Lasker}, {Le Goff}, {Le Guillou}, {Leauthaud}, {Levi}, {Li}, {Linder}, {Lodha}, {Magneville}, {Manera}, {Margala}, {Martini}, {Maus}, {McDonald}, {Medina-Varela}, {Meisner}, {Mena-Fern{\'a}ndez}, {Miquel}, {Moon}, {Moore}, {Moustakas}, {Mueller}, {Mu{\~n}oz-Guti{\'e}rrez}, {Myers}, {Nadathur}, {Napolitano}, {Neveux}, {Newman}, {Nguyen}, {Nie}, {Niz}, {Noriega}, {Padmanabhan}, {Paillas}, {Palanque-Delabrouille}, {Pan}, {Penmetsa}, {Percival}, {Pieri}, {Pinon}, {Poppett}, {Porredon}, {Prada}, {P{\'e}rez-Fern{\'a}ndez}, {P{\'e}rez-R{\`a}fols}, {Rabinowitz}, {Raichoor}, {Ram{\'\i}rez-P{\'e}rez}, {Ramirez-Solano}, {Rashkovetskyi}, {Ravoux}, {Rezaie}, {Rich}, {Rocher}, {Rockosi}, {Roe}, {Rosado-Marin}, {Ross}, {Rossi}, {Ruggeri}, {Ruhlmann-Kleider}, {Samushia}, {Sanchez}, {Saulder}, {Schlafly}, {Schlegel}, {Schubnell}, {Seo}, {Shafieloo}, {Sharples},
  {Silber}, {Slosar}, {Smith}, {Sprayberry}, {Tan}, {Tarl{\'e}}, {Taylor}, {Trusov}, {Ure{\~n}a-L{\'o}pez}, {Vaisakh}, {Valcin}, {Valdes}, {Vargas-Maga{\~n}a}, {Verde}, {Walther}, {Wang}, {Wang}, {Weaver}, {Weaverdyck}, {Wechsler}, {Weinberg}, {White}, {Yu}, {Yu}, {Yuan}, {Y{\`e}che}, {Zaborowski}, {Zarrouk}, {Zhang}, {Zhao}, {Zhao}, {Zhou} and {Zhuang}}]{DESI2025_BAO}
\bibinfo{author}{{Adame}, A.G.}, \bibinfo{author}{{Aguilar}, J.}, \bibinfo{author}{{Ahlen}, S.}, \bibinfo{author}{{Alam}, S.}, \bibinfo{author}{{Alexander}, D.M.}, \bibinfo{author}{{Alvarez}, M.}, \bibinfo{author}{{Alves}, O.}, \bibinfo{author}{{Anand}, A.}, \bibinfo{author}{{Andrade}, U.}, \bibinfo{author}{{Armengaud}, E.}, \bibinfo{author}{{Avila}, S.}, \bibinfo{author}{{Aviles}, A.}, \bibinfo{author}{{Awan}, H.}, \bibinfo{author}{{Bahr-Kalus}, B.}, \bibinfo{author}{{Bailey}, S.}, \bibinfo{author}{{Baltay}, C.}, \bibinfo{author}{{Bault}, A.}, \bibinfo{author}{{Behera}, J.}, \bibinfo{author}{{BenZvi}, S.}, \bibinfo{author}{{Bera}, A.}, \bibinfo{author}{{Beutler}, F.}, \bibinfo{author}{{Bianchi}, D.}, \bibinfo{author}{{Blake}, C.}, \bibinfo{author}{{Blum}, R.}, \bibinfo{author}{{Brieden}, S.}, \bibinfo{author}{{Brodzeller}, A.}, \bibinfo{author}{{Brooks}, D.}, \bibinfo{author}{{Buckley-Geer}, E.}, \bibinfo{author}{{Burtin}, E.}, \bibinfo{author}{{Calderon}, R.}, \bibinfo{author}{{Canning}, R.},
  \bibinfo{author}{{Carnero Rosell}, A.}, \bibinfo{author}{{Cereskaite}, R.}, \bibinfo{author}{{Cervantes-Cota}, J.L.}, \bibinfo{author}{{Chabanier}, S.}, \bibinfo{author}{{Chaussidon}, E.}, \bibinfo{author}{{Chaves-Montero}, J.}, \bibinfo{author}{{Chen}, S.}, \bibinfo{author}{{Chen}, X.}, \bibinfo{author}{{Claybaugh}, T.}, \bibinfo{author}{{Cole}, S.}, \bibinfo{author}{{Cuceu}, A.}, \bibinfo{author}{{Davis}, T.M.}, \bibinfo{author}{{Dawson}, K.}, \bibinfo{author}{{de la Macorra}, A.}, \bibinfo{author}{{de Mattia}, A.}, \bibinfo{author}{{Deiosso}, N.}, \bibinfo{author}{{Dey}, A.}, \bibinfo{author}{{Dey}, B.}, \bibinfo{author}{{Ding}, Z.}, \bibinfo{author}{{Doel}, P.}, \bibinfo{author}{{Edelstein}, J.}, \bibinfo{author}{{Eftekharzadeh}, S.}, \bibinfo{author}{{Eisenstein}, D.J.}, \bibinfo{author}{{Elliott}, A.}, \bibinfo{author}{{Fagrelius}, P.}, \bibinfo{author}{{Fanning}, K.}, \bibinfo{author}{{Ferraro}, S.}, \bibinfo{author}{{Ereza}, J.}, \bibinfo{author}{{Findlay}, N.}, \bibinfo{author}{{Flaugher}, B.},
  \bibinfo{author}{{Font-Ribera}, A.}, \bibinfo{author}{{Forero-S{\'a}nchez}, D.}, \bibinfo{author}{{Forero-Romero}, J.E.}, \bibinfo{author}{{Frenk}, C.S.}, \bibinfo{author}{{Garcia-Quintero}, C.}, \bibinfo{author}{{Gazta{\~n}aga}, E.}, \bibinfo{author}{{Gil-Mar{\'\i}n}, H.}, \bibinfo{author}{{Gontcho a Gontcho}, S.}, \bibinfo{author}{{Gonzalez-Morales}, A.X.}, \bibinfo{author}{{Gonzalez-Perez}, V.}, \bibinfo{author}{{Gordon}, C.}, \bibinfo{author}{{Green}, D.}, \bibinfo{author}{{Gruen}, D.}, \bibinfo{author}{{Gsponer}, R.}, \bibinfo{author}{{Gutierrez}, G.}, \bibinfo{author}{{Guy}, J.}, \bibinfo{author}{{Hadzhiyska}, B.}, \bibinfo{author}{{Hahn}, C.}, \bibinfo{author}{{Hanif}, M.M.S.}, \bibinfo{author}{{Herrera-Alcantar}, H.K.}, \bibinfo{author}{{Honscheid}, K.}, \bibinfo{author}{{Howlett}, C.}, \bibinfo{author}{{Huterer}, D.}, \bibinfo{author}{{Ir{\v{s}}i{\v{c}}}, V.}, \bibinfo{author}{{Ishak}, M.}, \bibinfo{author}{{Juneau}, S.}, \bibinfo{author}{{Kara{\c{c}}ayl{\i}}, N.G.}, \bibinfo{author}{{Kehoe}, R.},
  \bibinfo{author}{{Kent}, S.}, \bibinfo{author}{{Kirkby}, D.}, \bibinfo{author}{{Kremin}, A.}, \bibinfo{author}{{Krolewski}, A.}, \bibinfo{author}{{Lai}, Y.}, \bibinfo{author}{{Lan}, T.W.}, \bibinfo{author}{{Landriau}, M.}, \bibinfo{author}{{Lang}, D.}, \bibinfo{author}{{Lasker}, J.}, \bibinfo{author}{{Le Goff}, J.M.}, \bibinfo{author}{{Le Guillou}, L.}, \bibinfo{author}{{Leauthaud}, A.}, \bibinfo{author}{{Levi}, M.E.}, \bibinfo{author}{{Li}, T.S.}, \bibinfo{author}{{Linder}, E.}, \bibinfo{author}{{Lodha}, K.}, \bibinfo{author}{{Magneville}, C.}, \bibinfo{author}{{Manera}, M.}, \bibinfo{author}{{Margala}, D.}, \bibinfo{author}{{Martini}, P.}, \bibinfo{author}{{Maus}, M.}, \bibinfo{author}{{McDonald}, P.}, \bibinfo{author}{{Medina-Varela}, L.}, \bibinfo{author}{{Meisner}, A.}, \bibinfo{author}{{Mena-Fern{\'a}ndez}, J.}, \bibinfo{author}{{Miquel}, R.}, \bibinfo{author}{{Moon}, J.}, \bibinfo{author}{{Moore}, S.}, \bibinfo{author}{{Moustakas}, J.}, \bibinfo{author}{{Mueller}, E.},
  \bibinfo{author}{{Mu{\~n}oz-Guti{\'e}rrez}, A.}, \bibinfo{author}{{Myers}, A.D.}, \bibinfo{author}{{Nadathur}, S.}, \bibinfo{author}{{Napolitano}, L.}, \bibinfo{author}{{Neveux}, R.}, \bibinfo{author}{{Newman}, J.A.}, \bibinfo{author}{{Nguyen}, N.M.}, \bibinfo{author}{{Nie}, J.}, \bibinfo{author}{{Niz}, G.}, \bibinfo{author}{{Noriega}, H.E.}, \bibinfo{author}{{Padmanabhan}, N.}, \bibinfo{author}{{Paillas}, E.}, \bibinfo{author}{{Palanque-Delabrouille}, N.}, \bibinfo{author}{{Pan}, J.}, \bibinfo{author}{{Penmetsa}, S.}, \bibinfo{author}{{Percival}, W.J.}, \bibinfo{author}{{Pieri}, M.M.}, \bibinfo{author}{{Pinon}, M.}, \bibinfo{author}{{Poppett}, C.}, \bibinfo{author}{{Porredon}, A.}, \bibinfo{author}{{Prada}, F.}, \bibinfo{author}{{P{\'e}rez-Fern{\'a}ndez}, A.}, \bibinfo{author}{{P{\'e}rez-R{\`a}fols}, I.}, \bibinfo{author}{{Rabinowitz}, D.}, \bibinfo{author}{{Raichoor}, A.}, \bibinfo{author}{{Ram{\'\i}rez-P{\'e}rez}, C.}, \bibinfo{author}{{Ramirez-Solano}, S.}, \bibinfo{author}{{Rashkovetskyi}, M.},
  \bibinfo{author}{{Ravoux}, C.}, \bibinfo{author}{{Rezaie}, M.}, \bibinfo{author}{{Rich}, J.}, \bibinfo{author}{{Rocher}, A.}, \bibinfo{author}{{Rockosi}, C.}, \bibinfo{author}{{Roe}, N.A.}, \bibinfo{author}{{Rosado-Marin}, A.}, \bibinfo{author}{{Ross}, A.J.}, \bibinfo{author}{{Rossi}, G.}, \bibinfo{author}{{Ruggeri}, R.}, \bibinfo{author}{{Ruhlmann-Kleider}, V.}, \bibinfo{author}{{Samushia}, L.}, \bibinfo{author}{{Sanchez}, E.}, \bibinfo{author}{{Saulder}, C.}, \bibinfo{author}{{Schlafly}, E.F.}, \bibinfo{author}{{Schlegel}, D.}, \bibinfo{author}{{Schubnell}, M.}, \bibinfo{author}{{Seo}, H.}, \bibinfo{author}{{Shafieloo}, A.}, \bibinfo{author}{{Sharples}, R.}, \bibinfo{author}{{Silber}, J.}, \bibinfo{author}{{Slosar}, A.}, \bibinfo{author}{{Smith}, A.}, \bibinfo{author}{{Sprayberry}, D.}, \bibinfo{author}{{Tan}, T.}, \bibinfo{author}{{Tarl{\'e}}, G.}, \bibinfo{author}{{Taylor}, P.}, \bibinfo{author}{{Trusov}, S.}, \bibinfo{author}{{Ure{\~n}a-L{\'o}pez}, L.A.}, \bibinfo{author}{{Vaisakh}, R.},
  \bibinfo{author}{{Valcin}, D.}, \bibinfo{author}{{Valdes}, F.}, \bibinfo{author}{{Vargas-Maga{\~n}a}, M.}, \bibinfo{author}{{Verde}, L.}, \bibinfo{author}{{Walther}, M.}, \bibinfo{author}{{Wang}, B.}, \bibinfo{author}{{Wang}, M.S.}, \bibinfo{author}{{Weaver}, B.A.}, \bibinfo{author}{{Weaverdyck}, N.}, \bibinfo{author}{{Wechsler}, R.H.}, \bibinfo{author}{{Weinberg}, D.H.}, \bibinfo{author}{{White}, M.}, \bibinfo{author}{{Yu}, J.}, \bibinfo{author}{{Yu}, Y.}, \bibinfo{author}{{Yuan}, S.}, \bibinfo{author}{{Y{\`e}che}, C.}, \bibinfo{author}{{Zaborowski}, E.A.}, \bibinfo{author}{{Zarrouk}, P.}, \bibinfo{author}{{Zhang}, H.}, \bibinfo{author}{{Zhao}, C.}, \bibinfo{author}{{Zhao}, R.}, \bibinfo{author}{{Zhou}, R.}, \bibinfo{author}{{Zhuang}, T.}, \bibinfo{year}{2025}.
\newblock \bibinfo{title}{{DESI 2024 VI: cosmological constraints from the measurements of baryon acoustic oscillations}}.
\newblock \bibinfo{journal}{\jcap} \bibinfo{volume}{2025}, \bibinfo{pages}{021}.
\newblock \DOIprefix\doi{10.1088/1475-7516/2025/02/021}, \href{http://arxiv.org/abs/2404.03002}{{\tt arXiv:2404.03002}}.
\bibitem[{{Altmayer Pizzorno} and {Berger}(2024)}]{Pizzorno2024}
\bibinfo{author}{{Altmayer Pizzorno}, J.}, \bibinfo{author}{{Berger}, E.D.}, \bibinfo{year}{2024}.
\newblock \bibinfo{title}{{CoverUp: Effective High Coverage Test Generation for Python}}.
\newblock \bibinfo{journal}{arXiv e-prints} , \bibinfo{pages}{arXiv:2403.16218}\DOIprefix\doi{10.48550/arXiv.2403.16218}, \href{http://arxiv.org/abs/2403.16218}{{\tt arXiv:2403.16218}}.
\bibitem[{{Amendola} et~al.(1999){Amendola}, {Frieman} and {Waga}}]{Amendola1999}
\bibinfo{author}{{Amendola}, L.}, \bibinfo{author}{{Frieman}, J.A.}, \bibinfo{author}{{Waga}, I.}, \bibinfo{year}{1999}.
\newblock \bibinfo{title}{{Weak gravitational lensing by voids}}.
\newblock \bibinfo{journal}{\mnras} \bibinfo{volume}{309}, \bibinfo{pages}{465--473}.
\newblock \DOIprefix\doi{10.1046/j.1365-8711.1999.02841.x}, \href{http://arxiv.org/abs/astro-ph/9811458}{{\tt arXiv:astro-ph/9811458}}.
\bibitem[{{Bartelmann} and {Schneider}(2001)}]{Bartelmann2001}
\bibinfo{author}{{Bartelmann}, M.}, \bibinfo{author}{{Schneider}, P.}, \bibinfo{year}{2001}.
\newblock \bibinfo{title}{{Weak gravitational lensing}}.
\newblock \bibinfo{journal}{\physrep} \bibinfo{volume}{340}, \bibinfo{pages}{291--472}.
\newblock \DOIprefix\doi{10.1016/S0370-1573(00)00082-X}, \href{http://arxiv.org/abs/astro-ph/9912508}{{\tt arXiv:astro-ph/9912508}}.
\bibitem[{{Birrer} and {Amara}(2018)}]{Birrer2018}
\bibinfo{author}{{Birrer}, S.}, \bibinfo{author}{{Amara}, A.}, \bibinfo{year}{2018}.
\newblock \bibinfo{title}{{lenstronomy: Multi-purpose gravitational lens modelling software package}}.
\newblock \bibinfo{journal}{Physics of the Dark Universe} \bibinfo{volume}{22}, \bibinfo{pages}{189--201}.
\newblock \DOIprefix\doi{10.1016/j.dark.2018.11.002}, \href{http://arxiv.org/abs/1803.09746}{{\tt arXiv:1803.09746}}.
\bibitem[{{Birrer} et~al.(2015){Birrer}, {Amara} and {Refregier}}]{Birrer2015}
\bibinfo{author}{{Birrer}, S.}, \bibinfo{author}{{Amara}, A.}, \bibinfo{author}{{Refregier}, A.}, \bibinfo{year}{2015}.
\newblock \bibinfo{title}{{Gravitational Lens Modeling with Basis Sets}}.
\newblock \bibinfo{journal}{\apj} \bibinfo{volume}{813}, \bibinfo{pages}{102}.
\newblock \DOIprefix\doi{10.1088/0004-637X/813/2/102}, \href{http://arxiv.org/abs/1504.07629}{{\tt arXiv:1504.07629}}.
\bibitem[{Birrer et~al.(2021)Birrer, Shajib, Gilman, Galan, Aalbers, Millon, Morgan, Pagano, Park, Teodori, Tessore, Ueland, Van~de Vyvere, Wagner-Carena, Wempe, Yang, Ding, Schmidt, Sluse, Zhang and Amara}]{Birrer2021}
\bibinfo{author}{Birrer, S.}, \bibinfo{author}{Shajib, A.J.}, \bibinfo{author}{Gilman, D.}, \bibinfo{author}{Galan, A.}, \bibinfo{author}{Aalbers, J.}, \bibinfo{author}{Millon, M.}, \bibinfo{author}{Morgan, R.}, \bibinfo{author}{Pagano, G.}, \bibinfo{author}{Park, J.W.}, \bibinfo{author}{Teodori, L.}, \bibinfo{author}{Tessore, N.}, \bibinfo{author}{Ueland, M.}, \bibinfo{author}{Van~de Vyvere, L.}, \bibinfo{author}{Wagner-Carena, S.}, \bibinfo{author}{Wempe, E.}, \bibinfo{author}{Yang, L.}, \bibinfo{author}{Ding, X.}, \bibinfo{author}{Schmidt, T.}, \bibinfo{author}{Sluse, D.}, \bibinfo{author}{Zhang, M.}, \bibinfo{author}{Amara, A.}, \bibinfo{year}{2021}.
\newblock \bibinfo{title}{lenstronomy ii: A gravitational lensing software ecosystem}.
\newblock \bibinfo{journal}{Journal of Open Source Software} \bibinfo{volume}{6}, \bibinfo{pages}{3283}.
\newblock \URLprefix \url{https://doi.org/10.21105/joss.03283}, \DOIprefix\doi{10.21105/joss.03283}.
\bibitem[{{Buchert} and {Ehlers}(1993)}]{Buchert1993}
\bibinfo{author}{{Buchert}, T.}, \bibinfo{author}{{Ehlers}, J.}, \bibinfo{year}{1993}.
\newblock \bibinfo{title}{{Lagrangian theory of gravitational instability of Friedman-Lemaitre cosmologies -- second-order approach: an improved model for non-linear clustering}}.
\newblock \bibinfo{journal}{\mnras} \bibinfo{volume}{264}, \bibinfo{pages}{375--387}.
\newblock \DOIprefix\doi{10.1093/mnras/264.2.375}.
\bibitem[{{Cemri} et~al.(2025){Cemri}, {Pan}, {Yang}, {Agrawal}, {Chopra}, {Tiwari}, {Keutzer}, {Parameswaran}, {Klein}, {Ramchandran}, {Zaharia}, {Gonzalez} and {Stoica}}]{Cemri2025}
\bibinfo{author}{{Cemri}, M.}, \bibinfo{author}{{Pan}, M.Z.}, \bibinfo{author}{{Yang}, S.}, \bibinfo{author}{{Agrawal}, L.A.}, \bibinfo{author}{{Chopra}, B.}, \bibinfo{author}{{Tiwari}, R.}, \bibinfo{author}{{Keutzer}, K.}, \bibinfo{author}{{Parameswaran}, A.}, \bibinfo{author}{{Klein}, D.}, \bibinfo{author}{{Ramchandran}, K.}, \bibinfo{author}{{Zaharia}, M.}, \bibinfo{author}{{Gonzalez}, J.E.}, \bibinfo{author}{{Stoica}, I.}, \bibinfo{year}{2025}.
\newblock \bibinfo{title}{{Why Do Multi-Agent LLM Systems Fail?}}
\newblock \bibinfo{journal}{arXiv e-prints} , \bibinfo{pages}{arXiv:2503.13657}\DOIprefix\doi{10.48550/arXiv.2503.13657}, \href{http://arxiv.org/abs/2503.13657}{{\tt arXiv:2503.13657}}.
\bibitem[{{Chen} et~al.(2023){Chen}, {Hu}, {Zhi}, {Han}, {Deng} and {Yin}}]{Chen2023}
\bibinfo{author}{{Chen}, Y.}, \bibinfo{author}{{Hu}, Z.}, \bibinfo{author}{{Zhi}, C.}, \bibinfo{author}{{Han}, J.}, \bibinfo{author}{{Deng}, S.}, \bibinfo{author}{{Yin}, J.}, \bibinfo{year}{2023}.
\newblock \bibinfo{title}{{ChatUniTest: A Framework for LLM-Based Test Generation}}.
\newblock \bibinfo{journal}{arXiv e-prints} , \bibinfo{pages}{arXiv:2305.04764}\DOIprefix\doi{10.48550/arXiv.2305.04764}, \href{http://arxiv.org/abs/2305.04764}{{\tt arXiv:2305.04764}}.
\bibitem[{{Eisenstein} and {Hu}(1998)}]{Eisenstein1998}
\bibinfo{author}{{Eisenstein}, D.J.}, \bibinfo{author}{{Hu}, W.}, \bibinfo{year}{1998}.
\newblock \bibinfo{title}{{Baryonic Features in the Matter Transfer Function}}.
\newblock \bibinfo{journal}{\apj} \bibinfo{volume}{496}, \bibinfo{pages}{605--614}.
\newblock \DOIprefix\doi{10.1086/305424}, \href{http://arxiv.org/abs/astro-ph/9709112}{{\tt arXiv:astro-ph/9709112}}.
\bibitem[{{Hamaus} et~al.(2014){Hamaus}, {Sutter} and {Wandelt}}]{Hamaus2014}
\bibinfo{author}{{Hamaus}, N.}, \bibinfo{author}{{Sutter}, P.M.}, \bibinfo{author}{{Wandelt}, B.D.}, \bibinfo{year}{2014}.
\newblock \bibinfo{title}{{Universal Density Profile for Cosmic Voids}}.
\newblock \bibinfo{journal}{\prl} \bibinfo{volume}{112}, \bibinfo{pages}{251302}.
\newblock \DOIprefix\doi{10.1103/PhysRevLett.112.251302}, \href{http://arxiv.org/abs/1403.5499}{{\tt arXiv:1403.5499}}.
\bibitem[{{Hanwen Shen} and {Tamkin}(2026)}]{HanwenShen2026}
\bibinfo{author}{{Hanwen Shen}, J.}, \bibinfo{author}{{Tamkin}, A.}, \bibinfo{year}{2026}.
\newblock \bibinfo{title}{{How AI Impacts Skill Formation}}.
\newblock \bibinfo{journal}{arXiv e-prints} , \bibinfo{pages}{arXiv:2601.20245}\DOIprefix\doi{10.48550/arXiv.2601.20245}, \href{http://arxiv.org/abs/2601.20245}{{\tt arXiv:2601.20245}}.
\bibitem[{{Hockney} and {Eastwood}(1988)}]{Hockney1988}
\bibinfo{author}{{Hockney}, R.W.}, \bibinfo{author}{{Eastwood}, J.W.}, \bibinfo{year}{1988}.
\newblock \bibinfo{title}{{Computer simulation using particles}}.
\bibitem[{{Hogg}(2026)}]{Hogg2026}
\bibinfo{author}{{Hogg}, D.W.}, \bibinfo{year}{2026}.
\newblock \bibinfo{title}{{Why do we do astrophysics?}}
\newblock \bibinfo{journal}{arXiv e-prints} , \bibinfo{pages}{arXiv:2602.10181}\DOIprefix\doi{10.48550/arXiv.2602.10181}, \href{http://arxiv.org/abs/2602.10181}{{\tt arXiv:2602.10181}}.
\bibitem[{Jimenez et~al.(2024)Jimenez, Yang, Wettig, Yao, Pei, Press and Narasimhan}]{Jimenez2024}
\bibinfo{author}{Jimenez, C.E.}, \bibinfo{author}{Yang, J.}, \bibinfo{author}{Wettig, A.}, \bibinfo{author}{Yao, S.}, \bibinfo{author}{Pei, K.}, \bibinfo{author}{Press, O.}, \bibinfo{author}{Narasimhan, K.}, \bibinfo{year}{2024}.
\newblock \bibinfo{title}{{SWE}-bench: Can language models resolve real-world {GitHub} issues?}, in: \bibinfo{booktitle}{The Twelfth International Conference on Learning Representations (ICLR)}.
\newblock \URLprefix \url{https://openreview.net/forum?id=VTF8yNQM66}, \DOIprefix\doi{10.48550/arXiv.2310.06770}, \href{http://arxiv.org/abs/2310.06770}{{\tt arXiv:2310.06770}}. \bibinfo{note}{originally published as arXiv:2310.06770}.
\bibitem[{{Kayser} et~al.(1986){Kayser}, {Refsdal} and {Stabell}}]{Kayser1986}
\bibinfo{author}{{Kayser}, R.}, \bibinfo{author}{{Refsdal}, S.}, \bibinfo{author}{{Stabell}, R.}, \bibinfo{year}{1986}.
\newblock \bibinfo{title}{{Astrophysical applications of gravitational micro-lensing.}}
\newblock \bibinfo{journal}{\aap} \bibinfo{volume}{166}, \bibinfo{pages}{36--52}.
\bibitem[{{Krause} et~al.(2013){Krause}, {Chang}, {Dor{\'e}} and {Umetsu}}]{Krause2013}
\bibinfo{author}{{Krause}, E.}, \bibinfo{author}{{Chang}, T.C.}, \bibinfo{author}{{Dor{\'e}}, O.}, \bibinfo{author}{{Umetsu}, K.}, \bibinfo{year}{2013}.
\newblock \bibinfo{title}{{The Weight of Emptiness: The Gravitational Lensing Signal of Stacked Voids}}.
\newblock \bibinfo{journal}{\apjl} \bibinfo{volume}{762}, \bibinfo{pages}{L20}.
\newblock \DOIprefix\doi{10.1088/2041-8205/762/2/L20}, \href{http://arxiv.org/abs/1210.2446}{{\tt arXiv:1210.2446}}.
\bibitem[{{Laverick} et~al.(2024){Laverick}, {Surrao}, {Zubeldia}, {Bolliet}, {Cranmer}, {Lewis}, {Sherwin} and {Lesgourgues}}]{Laverick2024}
\bibinfo{author}{{Laverick}, A.}, \bibinfo{author}{{Surrao}, K.}, \bibinfo{author}{{Zubeldia}, I.}, \bibinfo{author}{{Bolliet}, B.}, \bibinfo{author}{{Cranmer}, M.}, \bibinfo{author}{{Lewis}, A.}, \bibinfo{author}{{Sherwin}, B.}, \bibinfo{author}{{Lesgourgues}, J.}, \bibinfo{year}{2024}.
\newblock \bibinfo{title}{{Multi-Agent System for Cosmological Parameter Analysis}}.
\newblock \bibinfo{journal}{arXiv e-prints} , \bibinfo{pages}{arXiv:2412.00431}\DOIprefix\doi{10.48550/arXiv.2412.00431}, \href{http://arxiv.org/abs/2412.00431}{{\tt arXiv:2412.00431}}.
\bibitem[{Mosqueira-Rey et~al.(2023)Mosqueira-Rey, Hern{\'a}ndez-Pereira, Alonso-R{\'i}os, Bobes-Bascar{\'a}n and Fern{\'a}ndez-Leal}]{MosqueiraRey2023}
\bibinfo{author}{Mosqueira-Rey, E.}, \bibinfo{author}{Hern{\'a}ndez-Pereira, E.}, \bibinfo{author}{Alonso-R{\'i}os, D.}, \bibinfo{author}{Bobes-Bascar{\'a}n, J.}, \bibinfo{author}{Fern{\'a}ndez-Leal, {\'A}.}, \bibinfo{year}{2023}.
\newblock \bibinfo{title}{Human-in-the-loop machine learning: a state of the art}.
\newblock \bibinfo{journal}{Artificial Intelligence Review} \bibinfo{volume}{56}, \bibinfo{pages}{3005--3054}.
\newblock \URLprefix \url{https://doi.org/10.1007/s10462-022-10246-w}, \DOIprefix\doi{10.1007/s10462-022-10246-w}.
\bibitem[{{Nadathur} et~al.(2020){Nadathur}, {Woodfinden}, {Percival}, {Aubert}, {Bautista}, {Dawson}, {Escoffier}, {Fromenteau}, {Gil-Mar{\'\i}n}, {Rich}, {Ross}, {Rossi}, {Maga{\~n}a}, {Brownstein} and {Schneider}}]{Nadathur2020}
\bibinfo{author}{{Nadathur}, S.}, \bibinfo{author}{{Woodfinden}, A.}, \bibinfo{author}{{Percival}, W.J.}, \bibinfo{author}{{Aubert}, M.}, \bibinfo{author}{{Bautista}, J.}, \bibinfo{author}{{Dawson}, K.}, \bibinfo{author}{{Escoffier}, S.}, \bibinfo{author}{{Fromenteau}, S.}, \bibinfo{author}{{Gil-Mar{\'\i}n}, H.}, \bibinfo{author}{{Rich}, J.}, \bibinfo{author}{{Ross}, A.J.}, \bibinfo{author}{{Rossi}, G.}, \bibinfo{author}{{Maga{\~n}a}, M.V.}, \bibinfo{author}{{Brownstein}, J.R.}, \bibinfo{author}{{Schneider}, D.P.}, \bibinfo{year}{2020}.
\newblock \bibinfo{title}{{The completed SDSS-IV extended baryon oscillation spectroscopic survey: geometry and growth from the anisotropic void-galaxy correlation function in the luminous red galaxy sample}}.
\newblock \bibinfo{journal}{\mnras} \bibinfo{volume}{499}, \bibinfo{pages}{4140--4157}.
\newblock \DOIprefix\doi{10.1093/mnras/staa3074}, \href{http://arxiv.org/abs/2008.06060}{{\tt arXiv:2008.06060}}.
\bibitem[{{Navarro} et~al.(1997){Navarro}, {Frenk} and {White}}]{Navarro1997}
\bibinfo{author}{{Navarro}, J.F.}, \bibinfo{author}{{Frenk}, C.S.}, \bibinfo{author}{{White}, S.D.M.}, \bibinfo{year}{1997}.
\newblock \bibinfo{title}{{A Universal Density Profile from Hierarchical Clustering}}.
\newblock \bibinfo{journal}{\apj} \bibinfo{volume}{490}, \bibinfo{pages}{493--508}.
\newblock \DOIprefix\doi{10.1086/304888}, \href{http://arxiv.org/abs/astro-ph/9611107}{{\tt arXiv:astro-ph/9611107}}.
\bibitem[{{Neyrinck}(2008)}]{Neyrinck2008}
\bibinfo{author}{{Neyrinck}, M.C.}, \bibinfo{year}{2008}.
\newblock \bibinfo{title}{{ZOBOV: a parameter-free void-finding algorithm}}.
\newblock \bibinfo{journal}{\mnras} \bibinfo{volume}{386}, \bibinfo{pages}{2101--2109}.
\newblock \DOIprefix\doi{10.1111/j.1365-2966.2008.13180.x}, \href{http://arxiv.org/abs/0712.3049}{{\tt arXiv:0712.3049}}.
\bibitem[{{Rampf} and {Hahn}(2021)}]{Rampf2021}
\bibinfo{author}{{Rampf}, C.}, \bibinfo{author}{{Hahn}, O.}, \bibinfo{year}{2021}.
\newblock \bibinfo{title}{{Shell-crossing in a {\ensuremath{\Lambda}}CDM Universe}}.
\newblock \bibinfo{journal}{\mnras} \bibinfo{volume}{501}, \bibinfo{pages}{L71--L75}.
\newblock \DOIprefix\doi{10.1093/mnrasl/slaa198}, \href{http://arxiv.org/abs/2010.12584}{{\tt arXiv:2010.12584}}.
\bibitem[{{Saeedi} et~al.(2025){Saeedi}, {Buckner}, {Aponte} and {Aghazadeh}}]{Saeedi2025}
\bibinfo{author}{{Saeedi}, D.}, \bibinfo{author}{{Buckner}, D.}, \bibinfo{author}{{Aponte}, J.C.}, \bibinfo{author}{{Aghazadeh}, A.}, \bibinfo{year}{2025}.
\newblock \bibinfo{title}{{AstroAgents: A Multi-Agent AI for Hypothesis Generation from Mass Spectrometry Data}}.
\newblock \bibinfo{journal}{arXiv e-prints} , \bibinfo{pages}{arXiv:2503.23170}\DOIprefix\doi{10.48550/arXiv.2503.23170}, \href{http://arxiv.org/abs/2503.23170}{{\tt arXiv:2503.23170}}.
\bibitem[{Schuster et~al.(2026a)Schuster, Bouchard, Zoubian and Frei}]{Schuster_CosmicWebExplorer_2026}
\bibinfo{author}{Schuster, N.}, \bibinfo{author}{Bouchard, S.}, \bibinfo{author}{Zoubian, J.}, \bibinfo{author}{Frei, D.}, \bibinfo{year}{2026}a.
\newblock \bibinfo{title}{Cosmic web explorer: Real-time large-scale structure in the browser}.
\newblock \URLprefix \url{https://github.com/nicosmo/cosmic_web_explorer}, \DOIprefix\doi{10.5281/zenodo.18915566}.
\bibitem[{{Schuster} et~al.(2023){Schuster}, {Hamaus}, {Dolag} and {Weller}}]{Schuster2023}
\bibinfo{author}{{Schuster}, N.}, \bibinfo{author}{{Hamaus}, N.}, \bibinfo{author}{{Dolag}, K.}, \bibinfo{author}{{Weller}, J.}, \bibinfo{year}{2023}.
\newblock \bibinfo{title}{{Why cosmic voids matter: nonlinear structure \& linear dynamics}}.
\newblock \bibinfo{journal}{\jcap} \bibinfo{volume}{2023}, \bibinfo{pages}{031}.
\newblock \DOIprefix\doi{10.1088/1475-7516/2023/05/031}, \href{http://arxiv.org/abs/2210.02457}{{\tt arXiv:2210.02457}}.
\bibitem[{Schuster et~al.(2026b)Schuster, Salcedo and Frei}]{schuster_lensing_2026}
\bibinfo{author}{Schuster, N.}, \bibinfo{author}{Salcedo, A.N.}, \bibinfo{author}{Frei, D.}, \bibinfo{year}{2026}b.
\newblock \bibinfo{title}{Visualizing gravitational lensing: v1.0.0}.
\newblock \URLprefix \url{https://github.com/nicosmo/lensing_visualization}, \DOIprefix\doi{10.5281/zenodo.18914869}.
\bibitem[{{Shao} et~al.(2024){Shao}, {Samuel}, {Jiang}, {Yang} and {Yang}}]{Shao2024}
\bibinfo{author}{{Shao}, Y.}, \bibinfo{author}{{Samuel}, V.}, \bibinfo{author}{{Jiang}, Y.}, \bibinfo{author}{{Yang}, J.}, \bibinfo{author}{{Yang}, D.}, \bibinfo{year}{2024}.
\newblock \bibinfo{title}{{Collaborative Gym: A Framework for Enabling and Evaluating Human-Agent Collaboration}}.
\newblock \bibinfo{journal}{arXiv e-prints} , \bibinfo{pages}{arXiv:2412.15701}\DOIprefix\doi{10.48550/arXiv.2412.15701}, \href{http://arxiv.org/abs/2412.15701}{{\tt arXiv:2412.15701}}.
\bibitem[{{Shojaee} et~al.(2025){Shojaee}, {Mirzadeh}, {Alizadeh}, {Horton}, {Bengio} and {Farajtabar}}]{Shojaee2025}
\bibinfo{author}{{Shojaee}, P.}, \bibinfo{author}{{Mirzadeh}, I.}, \bibinfo{author}{{Alizadeh}, K.}, \bibinfo{author}{{Horton}, M.}, \bibinfo{author}{{Bengio}, S.}, \bibinfo{author}{{Farajtabar}, M.}, \bibinfo{year}{2025}.
\newblock \bibinfo{title}{{The Illusion of Thinking: Understanding the Strengths and Limitations of Reasoning Models via the Lens of Problem Complexity}}.
\newblock \bibinfo{journal}{arXiv e-prints} , \bibinfo{pages}{arXiv:2506.06941}\DOIprefix\doi{10.48550/arXiv.2506.06941}, \href{http://arxiv.org/abs/2506.06941}{{\tt arXiv:2506.06941}}.
\bibitem[{{Song} et~al.(2026){Song}, {Han} and {Goodman}}]{Song2026}
\bibinfo{author}{{Song}, P.}, \bibinfo{author}{{Han}, P.}, \bibinfo{author}{{Goodman}, N.}, \bibinfo{year}{2026}.
\newblock \bibinfo{title}{{Large Language Model Reasoning Failures}}.
\newblock \bibinfo{journal}{arXiv e-prints} , \bibinfo{pages}{arXiv:2602.06176}\DOIprefix\doi{10.48550/arXiv.2602.06176}, \href{http://arxiv.org/abs/2602.06176}{{\tt arXiv:2602.06176}}.
\bibitem[{{Springel}(2005)}]{Springel2005}
\bibinfo{author}{{Springel}, V.}, \bibinfo{year}{2005}.
\newblock \bibinfo{title}{{The cosmological simulation code GADGET-2}}.
\newblock \bibinfo{journal}{\mnras} \bibinfo{volume}{364}, \bibinfo{pages}{1105--1134}.
\newblock \DOIprefix\doi{10.1111/j.1365-2966.2005.09655.x}, \href{http://arxiv.org/abs/astro-ph/0505010}{{\tt arXiv:astro-ph/0505010}}.
\bibitem[{{Starace} et~al.(2025){Starace}, {Jaffe}, {Sherburn}, {Aung}, {Shern Chan}, {Maksin}, {Dias}, {Mays}, {Kinsella}, {Thompson}, {Heidecke}, {Glaese} and {Patwardhan}}]{Starace2025}
\bibinfo{author}{{Starace}, G.}, \bibinfo{author}{{Jaffe}, O.}, \bibinfo{author}{{Sherburn}, D.}, \bibinfo{author}{{Aung}, J.}, \bibinfo{author}{{Shern Chan}, J.}, \bibinfo{author}{{Maksin}, L.}, \bibinfo{author}{{Dias}, R.}, \bibinfo{author}{{Mays}, E.}, \bibinfo{author}{{Kinsella}, B.}, \bibinfo{author}{{Thompson}, W.}, \bibinfo{author}{{Heidecke}, J.}, \bibinfo{author}{{Glaese}, A.}, \bibinfo{author}{{Patwardhan}, T.}, \bibinfo{year}{2025}.
\newblock \bibinfo{title}{{PaperBench: Evaluating AI's Ability to Replicate AI Research}}.
\newblock \bibinfo{journal}{arXiv e-prints} , \bibinfo{pages}{arXiv:2504.01848}\DOIprefix\doi{10.48550/arXiv.2504.01848}, \href{http://arxiv.org/abs/2504.01848}{{\tt arXiv:2504.01848}}.
\bibitem[{{Sutter} et~al.(2015){Sutter}, {Lavaux}, {Hamaus}, {Pisani}, {Wandelt}, {Warren}, {Villaescusa-Navarro}, {Zivick}, {Mao} and {Thompson}}]{Sutter2015}
\bibinfo{author}{{Sutter}, P.M.}, \bibinfo{author}{{Lavaux}, G.}, \bibinfo{author}{{Hamaus}, N.}, \bibinfo{author}{{Pisani}, A.}, \bibinfo{author}{{Wandelt}, B.D.}, \bibinfo{author}{{Warren}, M.}, \bibinfo{author}{{Villaescusa-Navarro}, F.}, \bibinfo{author}{{Zivick}, P.}, \bibinfo{author}{{Mao}, Q.}, \bibinfo{author}{{Thompson}, B.B.}, \bibinfo{year}{2015}.
\newblock \bibinfo{title}{{VIDE: The Void IDentification and Examination toolkit}}.
\newblock \bibinfo{journal}{Astronomy and Computing} \bibinfo{volume}{9}, \bibinfo{pages}{1--9}.
\newblock \DOIprefix\doi{10.1016/j.ascom.2014.10.002}, \href{http://arxiv.org/abs/1406.1191}{{\tt arXiv:1406.1191}}.
\bibitem[{Swanson et~al.(2024)Swanson, Wu, Bulaong, Pak and Zou}]{Swanson2024}
\bibinfo{author}{Swanson, K.}, \bibinfo{author}{Wu, W.}, \bibinfo{author}{Bulaong, N.L.}, \bibinfo{author}{Pak, J.E.}, \bibinfo{author}{Zou, J.}, \bibinfo{year}{2024}.
\newblock \bibinfo{title}{The virtual lab: Ai agents design new sars-cov-2 nanobodies with experimental validation}.
\newblock \bibinfo{journal}{bioRxiv} \URLprefix \url{https://doi.org/10.1101/2024.11.11.623004}, \DOIprefix\doi{10.1101/2024.11.11.623004}.
\bibitem[{{Tufano} et~al.(2024){Tufano}, {Agarwal}, {Jang}, {Zilouchian Moghaddam} and {Sundaresan}}]{Tufano2024}
\bibinfo{author}{{Tufano}, M.}, \bibinfo{author}{{Agarwal}, A.}, \bibinfo{author}{{Jang}, J.}, \bibinfo{author}{{Zilouchian Moghaddam}, R.}, \bibinfo{author}{{Sundaresan}, N.}, \bibinfo{year}{2024}.
\newblock \bibinfo{title}{{AutoDev: Automated AI-Driven Development}}.
\newblock \bibinfo{journal}{arXiv e-prints} , \bibinfo{pages}{arXiv:2403.08299}\DOIprefix\doi{10.48550/arXiv.2403.08299}, \href{http://arxiv.org/abs/2403.08299}{{\tt arXiv:2403.08299}}.
\bibitem[{{Villaescusa-Navarro} et~al.(2025){Villaescusa-Navarro}, {Bolliet}, {Villanueva-Domingo}, {Bayer}, {Acquah}, {Amancharla}, {Barzilay-Siegal}, {Bermejo}, {Bilodeau}, {C{\'a}rdenas Ram{\'\i}rez}, {Cranmer}, {Fran{\c{c}}a}, {Hahn}, {Jiang}, {Jimenez}, {Lee}, {Lerario}, {Mamun}, {Meier}, {Ojha}, {Protopapas}, {Roy}, {Spergel}, {Taranc{\'o}n-{\'A}lvarez}, {Tiwari}, {Viel}, {Wadekar}, {Wang}, {Wang}, {Xu}, {Yovel}, {Yue}, {Zhou}, {Zhu}, {Zou} and {Zubeldia}}]{VillaescusaNavarro2025}
\bibinfo{author}{{Villaescusa-Navarro}, F.}, \bibinfo{author}{{Bolliet}, B.}, \bibinfo{author}{{Villanueva-Domingo}, P.}, \bibinfo{author}{{Bayer}, A.E.}, \bibinfo{author}{{Acquah}, A.}, \bibinfo{author}{{Amancharla}, C.}, \bibinfo{author}{{Barzilay-Siegal}, A.}, \bibinfo{author}{{Bermejo}, P.}, \bibinfo{author}{{Bilodeau}, C.}, \bibinfo{author}{{C{\'a}rdenas Ram{\'\i}rez}, P.}, \bibinfo{author}{{Cranmer}, M.}, \bibinfo{author}{{Fran{\c{c}}a}, U.L.}, \bibinfo{author}{{Hahn}, C.}, \bibinfo{author}{{Jiang}, Y.F.}, \bibinfo{author}{{Jimenez}, R.}, \bibinfo{author}{{Lee}, J.Y.}, \bibinfo{author}{{Lerario}, A.}, \bibinfo{author}{{Mamun}, O.}, \bibinfo{author}{{Meier}, T.}, \bibinfo{author}{{Ojha}, A.A.}, \bibinfo{author}{{Protopapas}, P.}, \bibinfo{author}{{Roy}, S.}, \bibinfo{author}{{Spergel}, D.N.}, \bibinfo{author}{{Taranc{\'o}n-{\'A}lvarez}, P.}, \bibinfo{author}{{Tiwari}, U.}, \bibinfo{author}{{Viel}, M.}, \bibinfo{author}{{Wadekar}, D.}, \bibinfo{author}{{Wang}, C.}, \bibinfo{author}{{Wang}, B.Y.},
  \bibinfo{author}{{Xu}, L.}, \bibinfo{author}{{Yovel}, Y.}, \bibinfo{author}{{Yue}, S.}, \bibinfo{author}{{Zhou}, W.H.}, \bibinfo{author}{{Zhu}, Q.}, \bibinfo{author}{{Zou}, J.}, \bibinfo{author}{{Zubeldia}, {\'I}.}, \bibinfo{year}{2025}.
\newblock \bibinfo{title}{{The Denario project: Deep knowledge AI agents for scientific discovery}}.
\newblock \bibinfo{journal}{arXiv e-prints} , \bibinfo{pages}{arXiv:2510.26887}\DOIprefix\doi{10.48550/arXiv.2510.26887}, \href{http://arxiv.org/abs/2510.26887}{{\tt arXiv:2510.26887}}.
\bibitem[{{Wang} et~al.(2023){Wang}, {Hu}, {Lu}, {Zhu}, {Zhang}, {Subramaniam}, {Loomba}, {Zhang}, {Sun} and {Wang}}]{Wang2023}
\bibinfo{author}{{Wang}, X.}, \bibinfo{author}{{Hu}, Z.}, \bibinfo{author}{{Lu}, P.}, \bibinfo{author}{{Zhu}, Y.}, \bibinfo{author}{{Zhang}, J.}, \bibinfo{author}{{Subramaniam}, S.}, \bibinfo{author}{{Loomba}, A.R.}, \bibinfo{author}{{Zhang}, S.}, \bibinfo{author}{{Sun}, Y.}, \bibinfo{author}{{Wang}, W.}, \bibinfo{year}{2023}.
\newblock \bibinfo{title}{{SciBench: Evaluating College-Level Scientific Problem-Solving Abilities of Large Language Models}}.
\newblock \bibinfo{journal}{arXiv e-prints} , \bibinfo{pages}{arXiv:2307.10635}\DOIprefix\doi{10.48550/arXiv.2307.10635}, \href{http://arxiv.org/abs/2307.10635}{{\tt arXiv:2307.10635}}.
\bibitem[{{Weinberg} et~al.(2013){Weinberg}, {Mortonson}, {Eisenstein}, {Hirata}, {Riess} and {Rozo}}]{Weinberg2013}
\bibinfo{author}{{Weinberg}, D.H.}, \bibinfo{author}{{Mortonson}, M.J.}, \bibinfo{author}{{Eisenstein}, D.J.}, \bibinfo{author}{{Hirata}, C.}, \bibinfo{author}{{Riess}, A.G.}, \bibinfo{author}{{Rozo}, E.}, \bibinfo{year}{2013}.
\newblock \bibinfo{title}{{Observational probes of cosmic acceleration}}.
\newblock \bibinfo{journal}{\physrep} \bibinfo{volume}{530}, \bibinfo{pages}{87--255}.
\newblock \DOIprefix\doi{10.1016/j.physrep.2013.05.001}, \href{http://arxiv.org/abs/1201.2434}{{\tt arXiv:1201.2434}}.
\bibitem[{Zahavy(2026)}]{Zahavy2026}
\bibinfo{author}{Zahavy, T.}, \bibinfo{year}{2026}.
\newblock \bibinfo{title}{Llms can't jump}.
\newblock \URLprefix \url{https://philsci-archive.pitt.edu/28024/}.
\bibitem[{{Zel'dovich}(1970)}]{Zeldovich1970}
\bibinfo{author}{{Zel'dovich}, Y.B.}, \bibinfo{year}{1970}.
\newblock \bibinfo{title}{{Gravitational instability: An approximate theory for large density perturbations.}}
\newblock \bibinfo{journal}{\aap} \bibinfo{volume}{5}, \bibinfo{pages}{84--89}.

\end{thebibliography}

\end{document}